\documentclass[conference]{IEEEtran}
\usepackage[sort]{cite}
\usepackage{amsmath,amssymb,amsfonts}
\usepackage{graphicx}
\usepackage{textcomp}
\usepackage[hyphens]{url}
\usepackage{hyperref}
\usepackage{caption}
\usepackage{subcaption}
\usepackage[braket, qm]{qcircuit}
\usepackage{algpseudocode}
\usepackage{algorithm} 
\usepackage{bm}
\usepackage{siunitx}
\usepackage{balance}
\usepackage{xcolor}
\newboolean{showcomments}
\setboolean{showcomments}{true}
\ifthenelse{\boolean{showcomments}}
{ \newcommand{\mynote}[3]{
     \fbox{\bfseries\sffamily\scriptsize#1}
        {\small$\blacktriangleright$\textsf{\emph{\color{#3}{#2}}}$\blacktriangleleft$}}}
{ \newcommand{\mynote}[3]{}}
\definecolor{asparagus}{rgb}{0.53, 0.66, 0.42}

\usepackage{xcolor, soul}
\newcommand{\colorhl}[2][yellow]{#2}
\newcommand{\be}[1]{\textcolor{blue}{#1}}
\newcommand{\w}[1]{\textcolor{red}{#1}}

\newcommand{\cnot}{\texttt{CNOT}}
\newcommand{\sqcnot}{$\sqrt{\texttt{CNOT}}$}
\newcommand{\ncnot}[1]{$\sqrt[#1]{\texttt{CNOT}}$}
\newcommand{\sqiswap}{$\sqrt{\texttt{iSWAP}}$}
\newcommand{\niswap}[1]{$\sqrt[#1]{\text{\texttt{iSWAP}}}$}
\newcommand{\iswap}{\texttt{iSWAP}}
\newcommand{\swap}{\texttt{SWAP}}
\newcommand{\berk}{\texttt{B}}
\newcommand{\sqberk}{$\sqrt{\texttt{B}}$}
\newcommand{\ehaar}{$\mathbb{E}[\text{Haar}]$}

\def\BibTeX{{\rm B\kern-.05em{\sc i\kern-.025em b}\kern-.08em
    T\kern-.1667em\lower.7ex\hbox{E}\kern-.125emX}}

\title{Parallel Driving for Fast Quantum Computing Under Speed Limits}
\author{Evan McKinney$^\dagger$, Chao Zhou$^\mathsection$, Mingkang Xia$^\mathsection$, Michael Hatridge$^\mathsection$, Alex K. Jones$^\dagger$\\
\normalsize $^\dagger$Department of Electrical and Computer Engineering, $^\mathsection$Department of Physics and Astronomy\\
\normalsize University of Pittsburgh}

\begin{document}
\maketitle
\thispagestyle{plain}
\pagestyle{plain}


\begin{abstract}

Increasing quantum circuit fidelity requires an efficient instruction set to avoid errors from decoherence. The choice of a two-qubit (2Q) hardware basis gate depends on a quantum modulator's native Hamiltonian interactions and applied control drives. In this paper, we propose a collaborative design approach to select the best ratio of drive parameters that determine the \emph{best basis gate} for a particular modulator. This requires considering the \emph{theoretical computing power of the gate} along with the \emph{practical speed limit} of that gate, given the modulator drive parameters. The practical speed limit arises from the couplers' tolerance for strong driving when one or more pumps is applied,  for which some combinations can result in higher overall speed limits than others.  Moreover, as this 2Q basis gate is typically applied multiple times in succession, interleaved by 1Q gates applied directly to the qubits, the speed of the 1Q gates can become a limiting factor for the quantum circuit.  We propose a \emph{parallel-drive} approach that drives the modulator and qubits simultaneously, allowing a richer capability of the 2Q basis gate and in some cases for this 1Q drive time to be absorbed entirely into the 2Q operation. This allows increasingly short duration 2Q gates while mitigating a significant source of overhead in some quantum systems. On average, this approach can decrease circuit duration by 17.84\% and decrease infidelity for random 2Q gates by 10.5\% compared to the best basic 2Q gate, $\sqrt{\texttt{iSWAP}}$.

\end{abstract}

\section{Introduction}
\label{sec:intro}
Quantum Computers (QCs) leverage quantum superposition and entanglement which, unlike classical computers, allows the QC core computing element, or \textit{qubit}, to conceptually interact with all other qubits, simultaneously.  This provides the promise of solving problems that that are intractable for classical computers.  However, currently realized QCs are part of the Noisy Intermediate-Scale Quantum (NISQ) era.  NISQ machines with more than a hundred qubits can be readily created\cite{chow2021ibm}; however, the qubit interactions remain limited to small neighborhoods and these quantum operations---or quantum  \textit{gates}---have limited fidelity.  While these ``noisy'' quantum operations continue to improve, even the best gates typically do not exceed 99.9\% fidelity\cite{arute2019quantum, li2019realisation, zhao2020high, moskalenko2022high}.

Quantum interactions are realized through qubit-qubit coupling.  Coupling is possible when there is a physical connection between the qubits and is governed by a \textit{modulator}.  These modulators range from simple as capacitive couplings to more elaborate nonlinear circuits~\cite{chow2011simple,kandala2019error,chow2021ibm}. 
The major source of error in superconducting QC hardware, which is at the heart of machines by IBM and Google, comes from qubit decoherence.  Thus, continued improvements in quantum gate capabilities and speeds are required to increase feasible circuit depth.

A critical component to building better quantum circuits is to identify the \textit{best basis gate} that can be realized by the modulator.  The reason for selecting a single basis gate is that calibrating gates is an expensive process.  Otherwise, one could just calibrate every single possible gate, or at least each gate that is required for a particular quantum workload.  Also, gates must to be calibrated independently between each pair of qubits as each pair will require different parameters and frequencies to be addressed uniquely. Moreover, gate calibrations are finicky processes that drift over time, requiring periodic re-calibration~\cite{tornow2022minimum}.  Thus, a single gate calibration or determining multiple gate parameters as a function of the calibrated gate is necessary to make this process tractable~\cite{perez2021error}.


However, the metric for determining the best basis gate may not be clear.  A standard metric for determining the quality of a gate is to calculate its Haar score, which is its coverage of all possible gates among two qubits as represented in a 3D space by the Weyl Chamber~\cite{monodromy}.  While this is a good representation of the computational power of the gate, many quantum algorithms tend to be reduced to the \cnot{} family of gates to complete their computational work~\cite{nielsenchuang}. The remainder of the circuit typically requires the non-entangling \swap{} gate, primarily to move data on the machine's interconnection topology.  Thus, a basis gate that best optimizes these two operations of \cnot{} and \swap{} is also a useful metric.

In this work, we consider parametrically driven interactions, in which far off-resonant drives activate, usually non-rotating, terms in the coupler to power an effective two-body interaction between a pair of qubits~\cite{narla2016robust, leung_deterministic_2019, zhou2021modular}.  The drive amplitude(s) determine the gate speed, while also imprinting a phase on the resultant qubit-qubit coupling.  The most prominent examples of these drives are: (1) photon exchange/swapping/conversion between the modes, or \textit{conversion}, and (2) pair-photon creation/annihilation, also called two-mode squeezing, which produces \textit{gain} in parametric amplifiers. These drives can be applied in tandem in a single coupler~\cite{Sung2021}. 

One way to identify the set of basis gates enabled by this process is to explore the parametrically driven coupler in the form of its Hamiltonian expression and the drive parameters that can be used to tune different 2Q basis gates.  For instance, it is relatively straightforward to show that a Hamiltonian with gain and conversion terms that naturally implements the \iswap{} family of gates \textit{can also directly implement} the \cnot{} and \berk{} families of gates as well as other more exotic gates, all by changing the ratio of gain and conversion (see Section II below).  However, the pulse times to implement these gates is a function of the drive capacity of the modulator.

All of these parametrically-driven gates depend on an actuator/modulator that has an inherent \textit{speed limit}. The source of this speed limit is the physical limit of the drive capacity of the device for which over-driving the modulator can result in drive lines causing heating, instability in non-linear objects, ``bright-state''-ing, bifurcation, chaos, population leakage~\cite{theis2016simultaneous}, among others.  Speed limits are a fundamental property of parametric couplers,  otherwise there would be no fundamental limit to how often every gate can be made $10\times$ faster. Understanding the various physical mechanisms in determining these speed limits with the goal to improve them is an ongoing research effort both for parametrically driven qubit gates and the related field of parametrically driven amplifiers~\cite{liu2017josephson, planat2019understanding,ashhab2022speed}.

To further complicate the selection of the basis gate, the parametric drive terms, \textit{e.g.}, the gain and conversion terms, both contribute towards the speed limit, but combine in a non-linear way.  Thus, finding the fastest basis gate can become an optimization function of both the \textit{theoretical computing power} of the gate and the \textit{pulse time} of that gate due to the physical speed limit of that particular ratio of drive parameters.

\begin{figure}
    \centering
    \subfloat[Traditional Trajectory\label{fig:trad-decomp}]{    \includegraphics[width=.4\columnwidth]{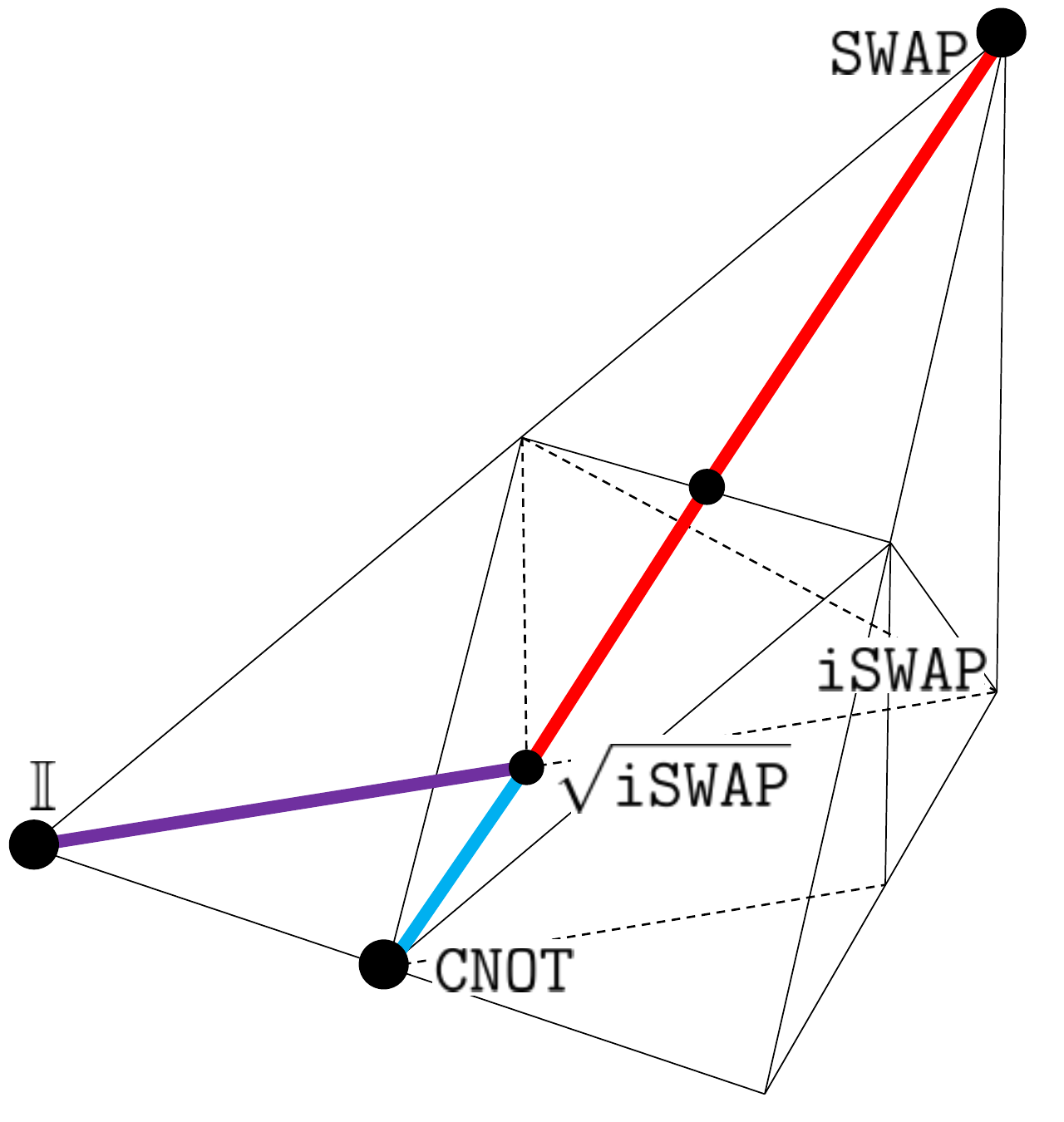}
    }
    \hfill
    \subfloat[Parallel-Driven Trajectory\label{fig:parallel-drive}]{
    \includegraphics[width=.4\columnwidth]{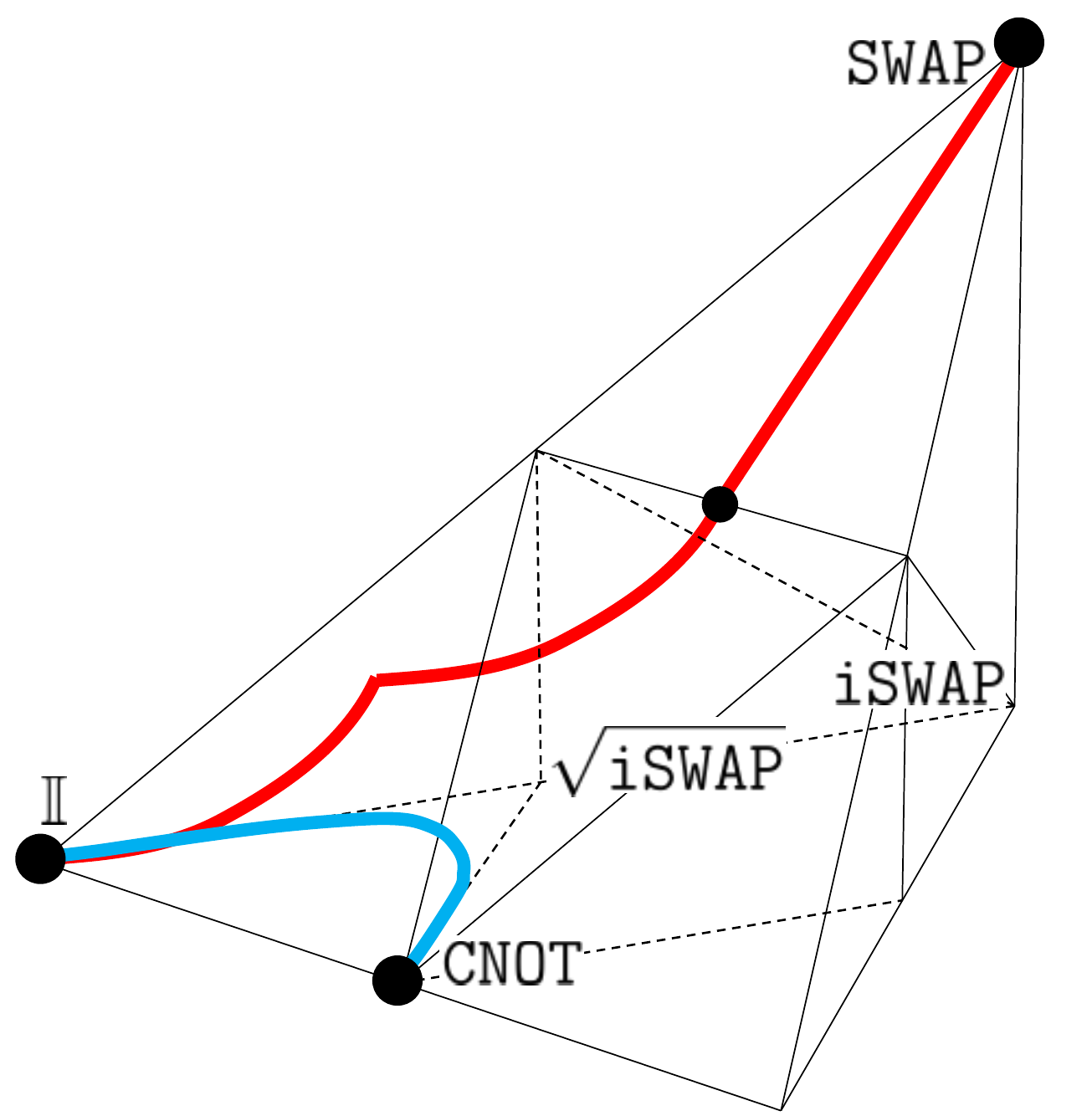}
    }
    \caption{Cartan Trajectories~\cite{Lin2022} for \cnot{} (blue) and \swap{} (red) using \sqiswap{} basis. \colorhl{The trajectories represent the total accumulated unitary transformation over time, beginning at Identity {$\mathbb{I}$} and ending at the target gate $U_T$. Black dots represent interleaved 1Q gates where orientation can be changed.}}
    \label{fig:cartan-traj}
    \vspace{-.1in}
\end{figure}

When conducting decomposition of a quantum circuit, traditionally, a template such that the selected 2Q basis gate is interspersed with 1Q gates (Fig.~\ref{fig:trad-decomp}). Conceptually, the 1Q gates orients the trajectory in a particular direction, then the 2Q basis gate traverses \textit{the Weyl chamber} to a new point.  The Weyl chamber is a 3D representation of the possible states of a pair of qubits.  Interestingly, to implement a \cnot{} or \swap{} functionality using \sqiswap{} the first leg in the Weyl chamber is identical, shown as purple. As the point of interest was not yet reached, the direction is re-oriented (1Q gates) before drawing the next line.  \cnot{} is reached in two steps, but the process repeats for \swap{} until the point is reached, which happens on the third step.  This process is akin to a car driving on established roads.  When on a road (path) the car must follow the road, but there are intersections at fixed points where the car can select new roads to follow.

In this paper we propose a collaboratively designed \textit{parallel-drive} technique that drives both the modulator and the qubits directly.  Intuitively, if the objective is the fastest path to the destination, stopping to steer adds delay. Instead, by driving both the qubits and the modulator simultaneously the straight paths become curves in the Weyl chamber, which is akin to turning while driving and breaking the restriction of driving on the established road.  This is shown  by allowing \cnot{} to be built from \iswap{} without intermediate 1Q gates and eliminating one set of interspersed 1Q gates in \swap{} (Fig.~\ref{fig:parallel-drive}). Moreover, parallel-drive allows the 2Q gate to increase in Haar volume and in some cases can eliminate the need for interspersed 1Q gates. More, it can expand the 2Q basis set without needing to calibrate new gates individually (assuming there is no cross talk between 1Q and 2Q gates), since there is still only one 2Q basis gate from which the others are created.

Thus, selecting high speed limit gate families and using increasingly short drive pulses while using parallel-drive to eliminate some 1Q gates can provide significant speed and fidelity improvements to implemented quantum circuits. In this paper, we make the following contributions:

\begin{itemize}
    \item We characterize simultaneous application of two basic parametric interactions to implement 2Q gates and articulate the various 2Q gate families that can be realized.
    \item We demonstrate that these modulators realize biased and potentially non-linear speed limits for different parameter drive ratios. 
    \item We observe that partially pulsed gates, \textit{e.g.}, \sqiswap, can be more efficient than the full pulse gate, \textit{e.g.}, \iswap.  However, when using smaller fractions of a gate, the overhead of the 1Q gates becomes more appreciable.
    \item We present a parallel-drive methodology to improve the agility of a basis gate by concurrently driving the modulator and the participating qubits.  We show that parallel-drive can improve the computing capability of a basis gate and provide the potential to remove interleaved 1Q gates in repeated application of 2Q gates.  
    \item Using these approaches we demonstrate an improved equivalency for \iswap{} and \cnot{} by using parallel-drive.
    \item We present a detailed study of using speed-limits and parallel-drive to reduce circuit delay for important quantum computing workloads.
\end{itemize}

In the next section we explore the basis gate design space from a modulator by exploring parameters of the Hamiltonian.

\section{Hamiltonian Design Space}
\label{sec:hamiltonian-design-space}
Fundamentally, quantum gates are unitary matrix operations, or unitaries, that act on quantum states. In general, 1Q and 2Q gates form the building blocks of quantum circuits~\cite{nielsenchuang}. A native quantum gate set, analogous to a classical computer's instruction set, defines which unitary operations are available to use on a machine. The available gates depends on the engineered Hamiltonian of the system, which is related to the unitary, described by Schr\"{o}dinger's equation, $U(t) = e^{-i \hat{H} t /\bar{h}}$. In superconducting QCs, parametric driving on a qubit-coupling mechanism provides control over the Hamiltonian to activate the desired unitary and corresponding gate.

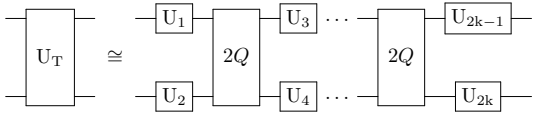
\begin{figure}[tbp]
\vspace{-.1in}
    \centering
\begin{equation*}
\resizebox{.8\columnwidth}{!}{
\Qcircuit @C=1.0em @R=0.4em @!R {
	 	& \multigate{2}{\mathrm{U_T}} & \qw & & & \gate{\mathrm{U_1}} & \multigate{2}{2Q} & \gate{\mathrm{U_3}} & \dots & & \multigate{2}{2Q} & \gate{\mathrm{U_{2k-1}}} & \qw
	 	\\
	 	& & & \push{\rule{0em}{0em}\!\!\!\! \cong \!\!\!\!\rule{0em}{0em}}
	 	\\
	 	& \ghost{\mathrm{U_T}} & \qw & & & \gate{\mathrm{U_2}} & \ghost{2Q} & \gate{\mathrm{U_4}} & \dots & & \ghost{2Q} & \gate{\mathrm{U_{2k}}} &\qw\\
}}
\end{equation*}
    \caption{Generic decomposition 2Q unitary $\leftarrow$ 2Q basis gate.}
    \label{fig:decomp-circuit}
    \vspace{-.15in}
\end{figure}

Using Cartan's KAK decomposition~\cite{khaneja2000cartan, tucci2005introduction}, an arbitrary 2Q gate can be built from repeated applications of a universal 2Q basis gate with interleaved 1Q gates (Fig.~\ref{fig:decomp-circuit}). Simple techniques for gate decomposition use this interleaving template and via an exact analytical solution~\cite{kalajdzievski2021exact, huang2021towards}, or an approximate numerical optimizer~\cite{rakyta2022approaching,smith2021leap, madden2022best}, find a solution to the 1Q gates for a variable number of repetitions. We refer to a \textit{basis template}, as a quantum circuit that interleaves the basis gate $K$ times. To perform decomposition, the template is instantiated with the sufficient size $K$.

Crucially, the proper selection of basis gate determines the overall complexity of the transpiled quantum algorithm, as different basis gates may require comparatively larger or smaller $K$ in decomposition. Moreover, each basis gate has a latency depending on the system's physical interactions. \textbf{For this reason, characterizing the set of candidate basis gates requires reasoning about both their decomposition efficiency, $K[U_B]$, as well as their hardware latency, $D[U_B]$.}

As discussed in Section~\ref{sec:intro}, the set of possible 2Q gates is represented geometrically by the Weyl Chamber~\cite{makhlin2002nonlocal, balakrishnan2009characterizing, zhang2003geometric}, where locally-equivalent 2Q gates, differing only by 1Q gates, are mapped to the same coordinate. 
This comes from the assumption that any \textit{locally equivalent} gate for a particular 2Q gate has the same entangling power and decomposition efficiency.   For example, \texttt{CZ} and \texttt{CX}/\cnot{} can be considered the same equivalent gates in this context.  Also, the unitary conjugates are reflected over the x-axis, which is like executing the gate backwards, so essentially it is only necessary to plot gates on the left side of the chamber. In this work, references to a 2Q basis gate may refer generally to the set of locally equivalent gates with matched computational power; however, references to 2Q target gates include the additional 1Q costs of local basis translation required for algorithm correctness.

To reason about decomposition, we plot a gate's coverage volume, which are regions that span all gates buildable by a template. We also use the $\in$ notation when referring to a template, which means the 1Q gates are free variables while the 2Q gates are fixed. The use of monodromy polytopes~\cite{monodromy} analytically creates the coverage sets, so we can reason about spanning volumes of $K$ gate applications, decide if a gate is contained in a template, and output its weighted volume.

\subsection{Flexible Realization of Gates with Parametric Couplings}

While certain classes of gate interactions, such as cross-resonance, rely on  linear coupling modes between superconducting qubits~\cite{rigetti2010fully, malekakhlagh2020first}, there is increasingly widespread use of \textit{parametric} or \textit{tunable} couplers among of groups of two or more qubits, which can be transmons, flux qubits, or high-Q cavities~\cite{Sung2021,Lin2022,Roy2022,Weiss2022,zhou2021modular}. In these systems, the coupler's nonlinearity is driven with external flux and/or microwave fields to create a wide array of potential gates.  Two well-known families of interactions used in parametric amplification~\cite{Roy2016} are photon exchange and two-mode squeezing/gain.  Photon exchange is produced by driving the coupler to provide the energy source/sink to exhange excitations between the qubits.  This can be realized, for instance, in a third-order coupler driven at the qubits' difference frequency.  

Two-mode squeezing/gain is produced by driving to provide the energy required for pair production/annihilation in the qubits.   In third-order coupling this requires driving at the qubits' sum frequency.  Surprisingly, both approaches naturally realize \iswap{} gates among two-level or anharmonic qubits.  Further, couplings which link to states outside the computational basis that create state dependent phase accumulation (\textit{e.g.}, difference driving between $\ket{11}$ and $\ket{20}$)  could be used to produce \texttt{CZ} gates, all in the same system. 

These couplers can be driven to produce a wide variety of 2Q gates, especially those in the Weyl chambers' floor, which are just combinations of simultaneous gain and conversion driving\footnote{Note, careful attention must be paid to the nonlinearity and encoding of the qubit states being used.  For instance, the same parametric interaction among qubits realized as transmons produces different gates than high-Q cavities, for which the latter produces Fock states.}. We can write such a combination Eq.~\ref{eq:conversion-gain},
\begin{equation}
\label{eq:conversion-gain}
\hat{H} = g_c (e^{i \phi_c} a^\dag b + e^{-i \phi_c}a b^\dag) + g_g (e^{i \phi_g}ab + e^{-i \phi_g}a^\dag b^\dag),
\end{equation}
where $g_c,g_g$ and $\phi_c,\phi_g$ represent the pump-controlled amplitude and phase, respectively, such that $g_c,\phi_c$ result from difference/conversion driving and $g_g,\phi_g$ are from sum/gain driving. 
Each choice of control parameters yields a continuum of gates. To illustrate the flexibility of jointly driving multiple interactions simultaneously, the case with both couplings are non-zero strength with both pump phases set to zero arrives at the following unitary:
\begin{equation}
\label{eq:unitary}
U(t) =  \begin{bmatrix}
        \cos{\theta_g} & 0 & 0 & -i\sin{\theta_g}\\
        0 & \cos{\theta_c} & -i\sin{\theta_c} & 0\\
        0 & -i\sin{\theta_c} & \cos{\theta_c} & 0\\
        -i\sin{\theta_g} & 0 & 0 & \cos{\theta_g}
    \end{bmatrix}\\
\end{equation}
such that $\theta_c = g_c t, \theta_g = g_g t$, where $t$ is the driving time.

By varying the interaction strengths $g_c$ and $g_g$ at a fixed $t=1$.  The \iswap{} gate in this language is given by setting $\theta_c$ or $\theta_g$ to $\frac{\pi}{2}$, yielding Eq.~\ref{eq:gain-or-conv-iswap}.
\begin{equation}
    \hat{H} = \frac{\pi}{2} (a^\dag b + a b^\dag) \text{ or } \hat{H} = \frac{\pi}{2} (ab + a^\dag b^\dag),
\label{eq:gain-or-conv-iswap}
\end{equation}
while the \cnot{} gate can be realized by setting $\theta_c=\theta_g=\frac{\pi}{2}$, yielding Eq.~\ref{eq:gain-eq-conv-cnot}.
\begin{equation}
    \hat{H} = \frac{\pi}{4} (a^\dag b + a b^\dag) + \frac{\pi}{4} (ab + a^\dag b^\dag).
\label{eq:gain-eq-conv-cnot}
    \vspace{-.1in}
\end{equation}
\begin{figure*}
    \subfloat[Set of gates natively produced by conversion and gain parametric driving. The color bar indicates $\theta_g + \theta_c$, normalized by $\pi/2$.] {
    \includegraphics[height=1.5in]{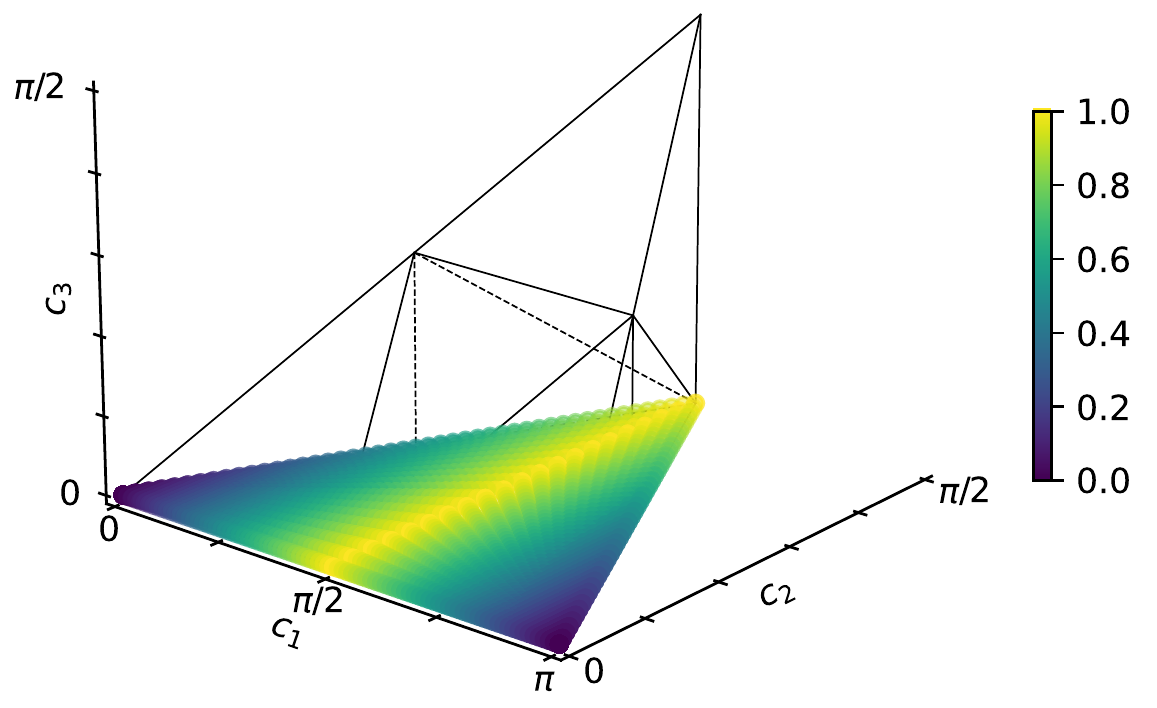}       \label{fig:possible_basis}
    }
    \hfill
    \subfloat[Frequency of gates from set of 16-qubit benchmarks transpiled onto an $4\times4$ square lattice topology.] {
    \includegraphics[height=1.5in]{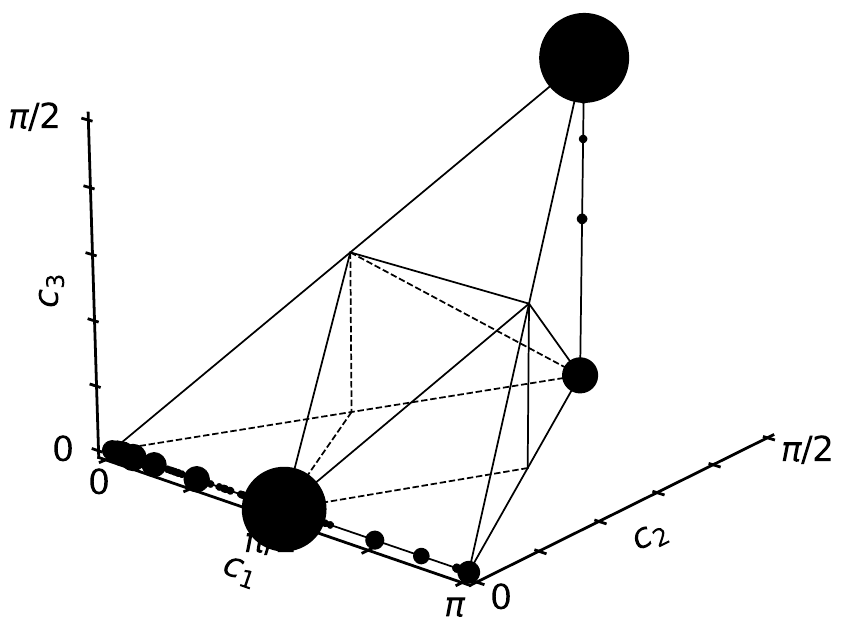}   
    \label{fig:basketball}
    }
    \hfill
    \subfloat[Demonstration of limitation of gain and conversion coefficients ($g_g$ and $g_c$) when both processes are turned on.]
    {
    \includegraphics[height=1.5in]{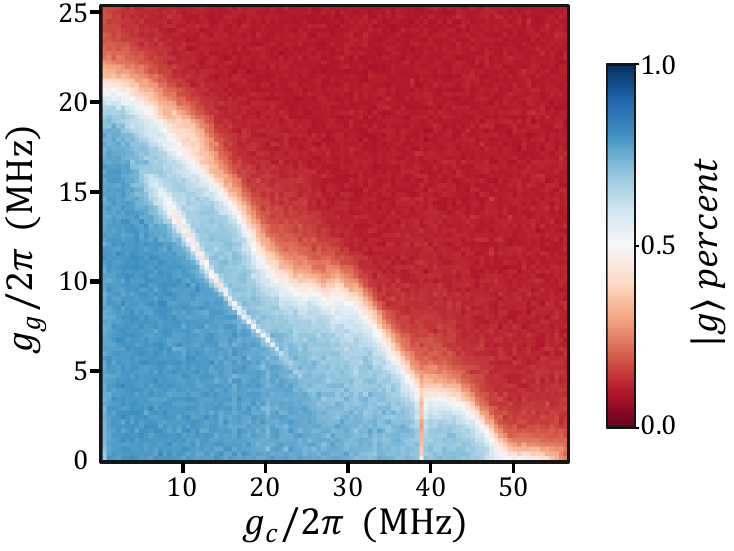}   
    \label{fig:snail_death}
    }
    \caption{Analysis of basis gate choice including, gate range and timing, gate usage by application, and impact of drive ratio. }
    \label{fig:pulse-times}
    \vspace{-.1in}
\end{figure*}

There is a continuous set of possible unitary operators that can be naturally realized by this Hamiltonian.  By visualizing this in the Weyl Chamber (Fig.~\ref{fig:possible_basis}) these two points of interest appear at both ends of the yellow band, with the \iswap{} at the tip and \cnot{} along the baseline at the $\frac{\pi}{2}$ point. In fact, the theoretical power of this Hamiltonian covers the entire base plane of the Weyl chamber with different points reachable in different ratios of $\theta_g$ and $\theta_c$ and total 
angle $\theta_g+\theta_c$. 

A vital question, then, is which combination of drives yields the best gate?  There are several important factors for selecting this gate such as the decompositional efficiency of the gate and the pulse time of the gate.  There is evidence that fractional pulse duration gates can be more efficient (\textit{e.g.}, \sqiswap{} vs \iswap{}), further reducing pulse time. In the next section we discuss methods to evaluate this decomposition efficiency. 




\subsection{Gate Score Methodology}
In order to optimize the choice of control parameters, it is first necessary to reason about the unitaries' decomposition efficiency. 
We compare two methodologies to quantify the decomposition efficiency of a gate: uniform gate distribution and algorithm-sampled distributions.
While decomposition determines the number of iterations required of a basis gate to realize a target unitary $U_T$, recall from Fig.~\ref{fig:possible_basis}, different realizable gates from the Hamiltonian require different pulse times.  To represent both aspects of a gate we define $K_{U_B}[U_T]$ as the number of basis gates ($U_B$) to build the target ($U_T$) and $D_{U_B}[U_T]$ as the normalized duration to build a target using the basis. The expectation $\mathbb{E}$ operator signifies the cost averaged over the random Haar distribution.

\begin{figure}[tbp]
\vspace{-.15in}
    \centering
    \subfloat[\iswap]{
    \includegraphics[width=.225\textwidth]{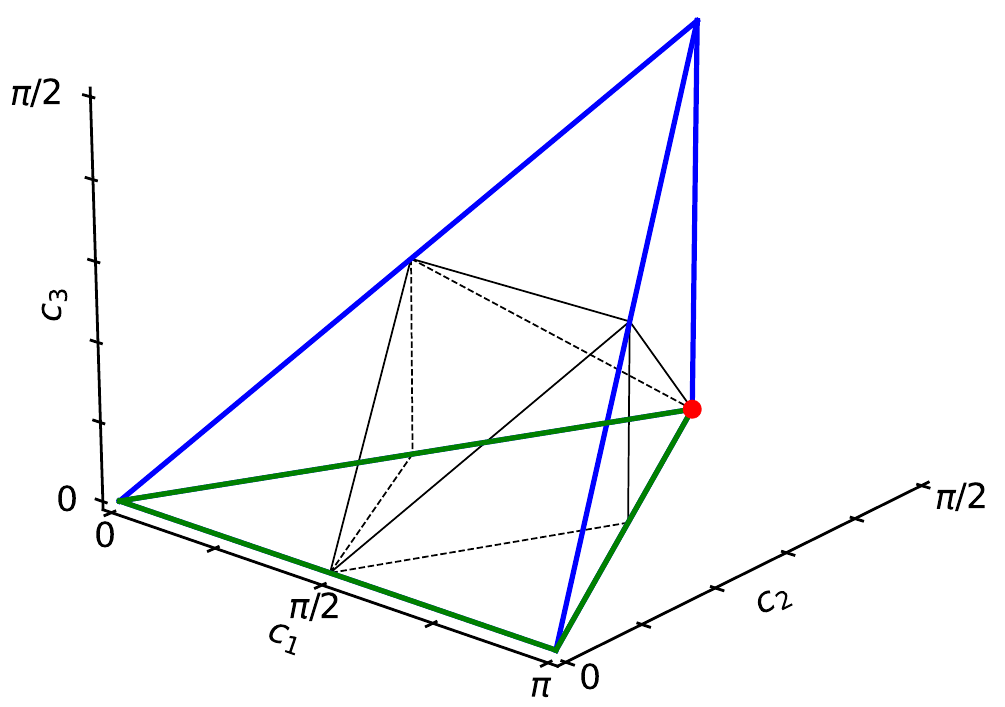}
    \label{fig:iswap_coverage}}
    \subfloat[\sqiswap]{
    \includegraphics[width=.225\textwidth]{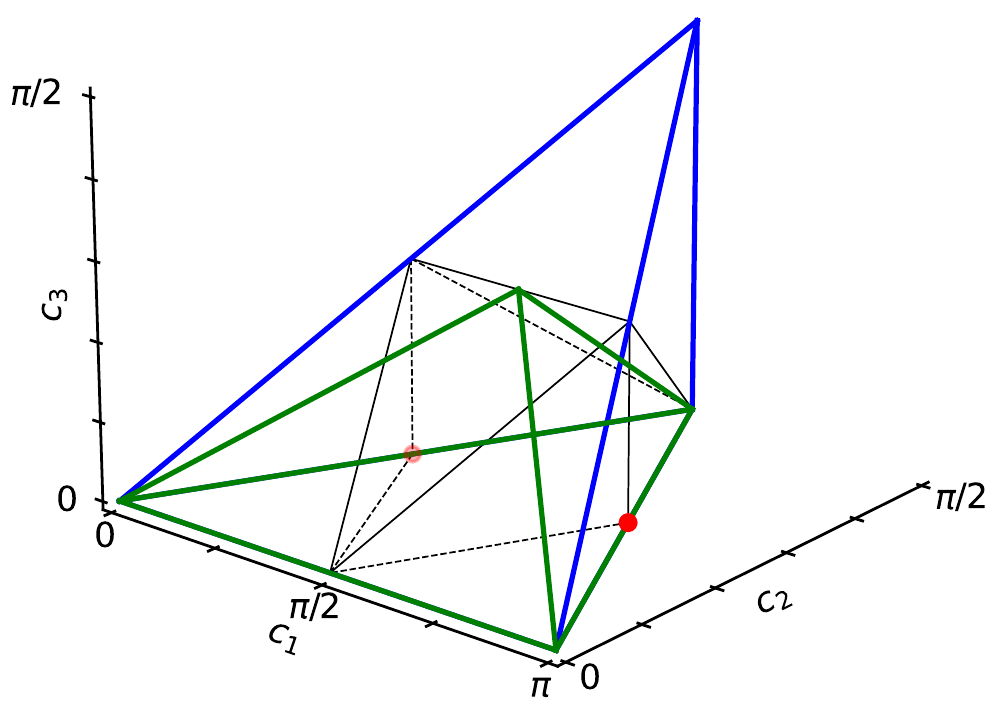}
    \label{fig:sqiswap_coverage}}
    \\
    \subfloat[\cnot]{
    \includegraphics[width=.225\textwidth]{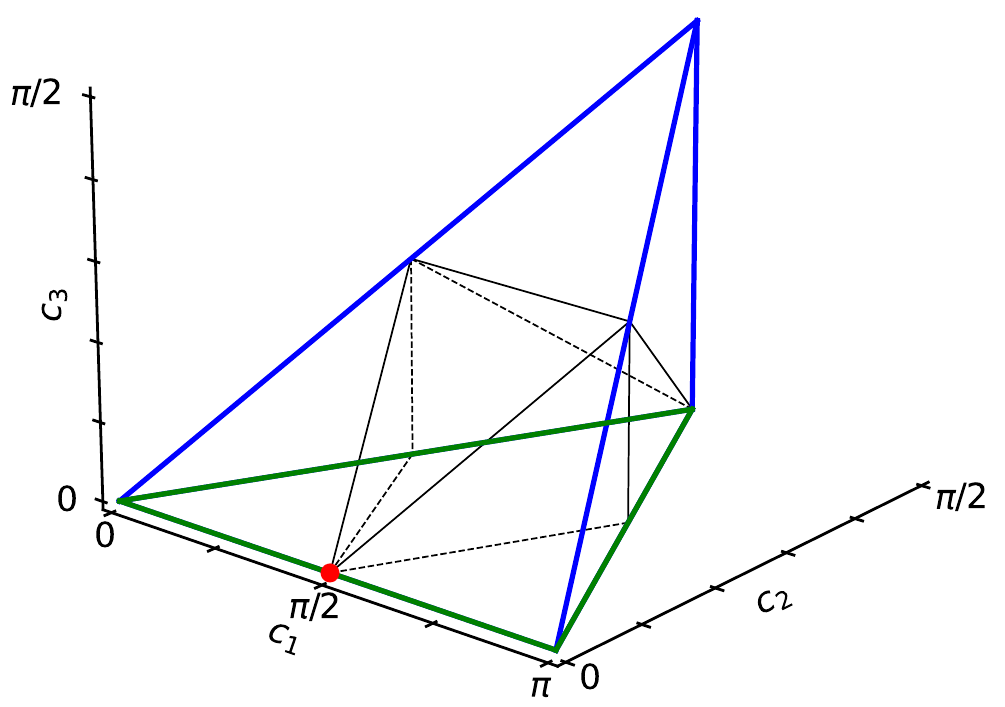}
    \label{fig:cnot_coverage}}
    \subfloat[\sqcnot]{
    \includegraphics[width=.225\textwidth]{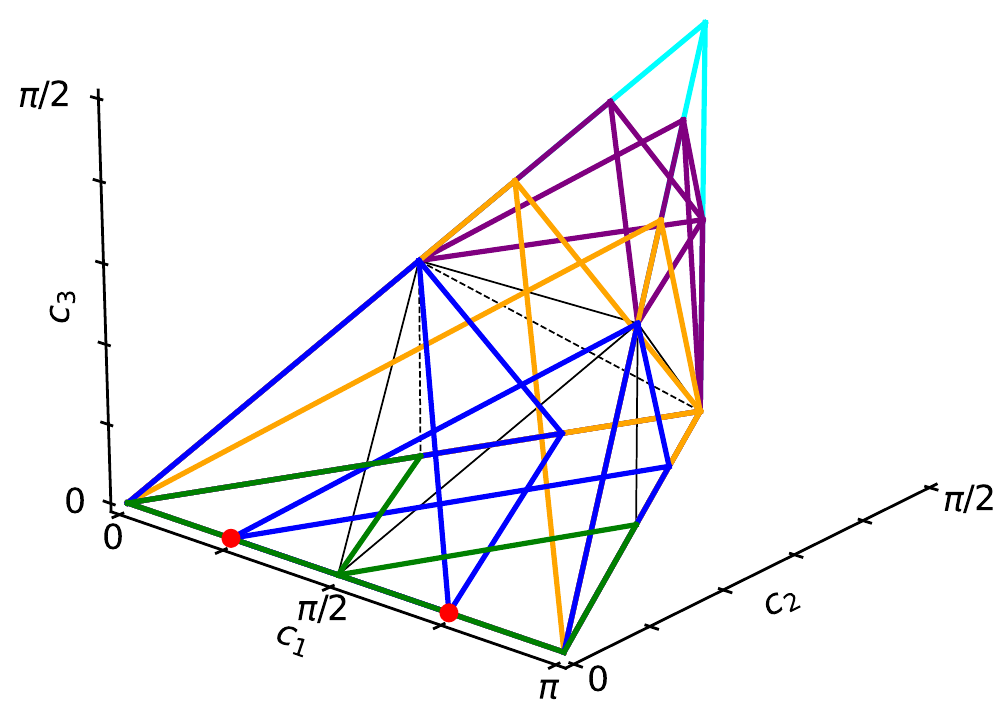}
    \label{fig:sqcnot_coverage}}
    \\
    \subfloat[\berk]{
    \includegraphics[width=.225\textwidth]{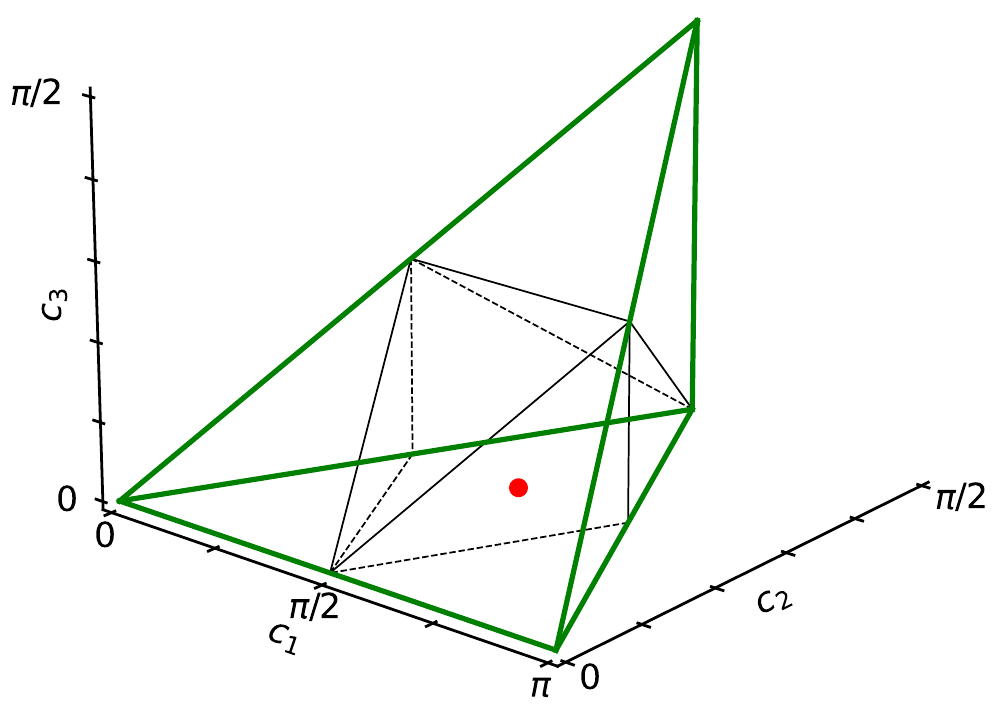}
    \label{fig:b_coverage}}
    \subfloat[\sqberk]{
    \includegraphics[width=.225\textwidth]{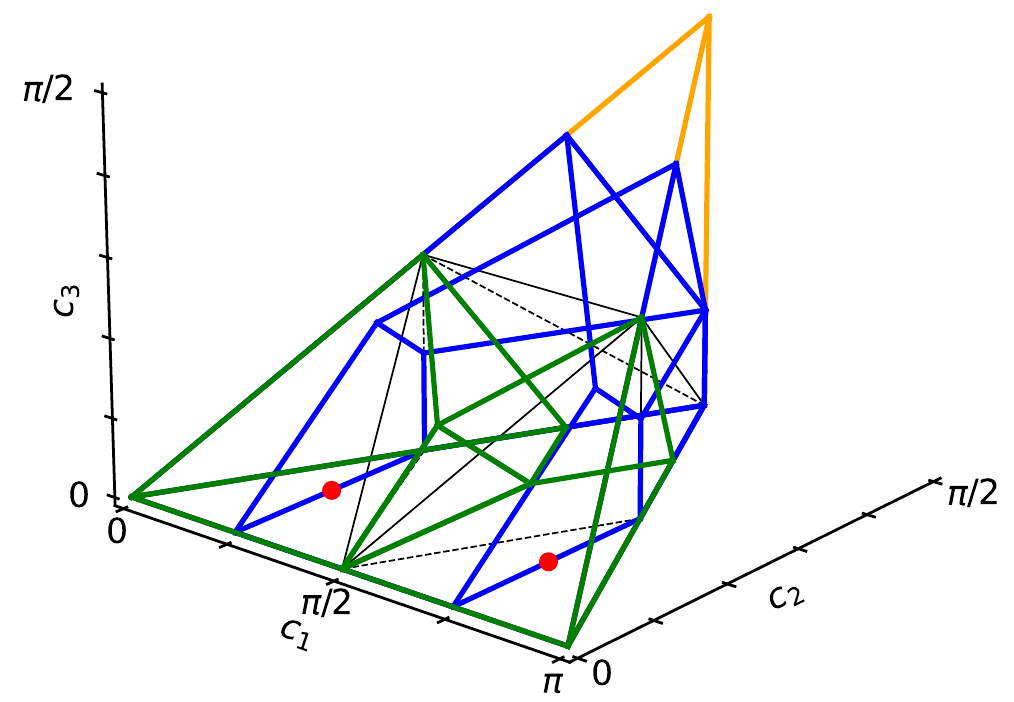}
    \label{fig:sqb_coverage}}
    \caption{Gate Coverage Sets. \textcolor{red}{red: $k=1$}, \textcolor{green}{green: $k=2$}, \textcolor{blue}{blue: $k=3$}, \textcolor{orange}{orange: $k=4$}, \textcolor{violet}{purple: $k=5$}, \textcolor{cyan}{cyan: $k=6$}}    \label{fig:coverage}
    \vspace{-.25in}
\end{figure}

The Haar measure~\cite{watts2013metric}, is used to construct a uniform distribution of 2Q gates. Conceptually, it is a density function inside the Weyl Chamber which weights the perfect entangler interior region more heavily than the exterior $\mathbb{I}$ (identity) and \texttt{SWAP} vertices.  It is used to build a Haar score, a common metric to quantify the decomposition power of a basis set.  The Haar score is the expected number of gates ($K$) to generate Haar random 2Q gates. In other words, it is a volume-weighted average over the basis template's spanning regions 
to achieve full Weyl Chamber coverage. This is demonstrated in  Fig.~\ref{fig:coverage}, by plotting the $K$-template spanning region for some popular 2Q gates.  The \iswap{} gate (Fig.~\ref{fig:iswap_coverage}) can reach the bottom plane in $k=2$ and the entire volume in $k=3$.  The \sqiswap{} gate (Fig.~\ref{fig:sqiswap_coverage}) actually has better coverage at $k=2$ with a shorter pulse time.  The popular \cnot{} (Fig.~\ref{fig:cnot_coverage}) has similar coverage behavior as \iswap{}, which is reasonable as both are Clifford gates. The \berk{} gate (Fig.~\ref{fig:b_coverage}) minimizes \ehaar{} because it spans the entirety of the region in $k=2$ (green), whereas \sqcnot{} (Fig.~\ref{fig:sqcnot_coverage}) does not completely span the chamber until six steps ($k=6$, yellow).

However, the \ehaar{} score fails to capture that in practice, gates are not uniformly distributed. Algorithms are written primarily using \texttt{CPhase} gates, implemented by \textit{controlled} unitaries, analytically decomposed into 2Q \cnot{} gates.  The reason why \texttt{CPhase} gates are ubiquitous in algorithm design may be explained by the Quantum Singular Value Transform (QSVT), a key subroutine of Grover's Search, Phase Estimation, and Hamiltonian Simulation circuits~\cite{martyn2021grand}.  The QSVT subroutine encodes an operator $\bm{A}$ as a block inside a larger unitary matrix, $\bm{U}$. When $\bm{A}$ is a unitary matrix, then $\bm{U}$ becomes a controlled $\bm{A}$ operator, which naturally decomposes into \texttt{CPhase} 2Q gates~\cite{nielsenchuang}. 
Even algorithms which use different subroutines, such as Quantum Approximate Optimization Algorithm (QAOA), still rely on the \cnot{} for their own reasons. In QAOA, the Hamiltonian cost function maps to states that are diagonal in the computational basis, such that the canonical expansion is into $\texttt{ZZ}$ gates~\cite{farhi2014quantum}. Simply, creating new quantum algorithms is such a difficult task, most known algorithms are variations of the same subroutine, which happens to use controlled-gate operators~\cite{shor2003haven}.

Moreover, qubit connectivity topologies necessitate data movement via \swap{} gates.  Due to the limited connection topologies of NISQ superconducting QCs of square lattice and heavy hex, \swap{} gates are required to move data into qubits in the same neighborhood. It has been shown that these gates can dominate transpiled gate counts~\cite{gokhale2021faster, mckinney}.

The frequency of \swap{} gates naturally depends on the coupling topology. For simplicity, we consider a 4x4 square lattice topology as the target coupling map. For a representative set of quantum benchmarks including QFT, QAOA, Adder, Multiplier, GHZ, Hidden Linear Function, and VQE, but excluding the special case of Quantum Volume, the workloads were mapped to this topology using the Qiskit v0.20.2 transpiler with \texttt{-O3} (optimization level 3), inducing the necessary \swap{}s. The results are displayed in a ``shot-chart'' that increases the size of the plotted gates relative to its frequency in the workloads, as shown in Fig.~\ref{fig:basketball}.  From this experiment, the most frequent targeted gates are \swap{} followed by \cnot{}, with \iswap{} as a more distant third.  Interestingly, there is a significant usage of \cnot{} \textit{family} gates, which show up along the Weyl chamber baseline.

Thus, an alternative gate scoring function introduces $V(U_B)$, which weights the decomposition cost of target, $U_T$, using the basis gate duration, $D_{U_B}[U_T]$, by the frequency of the target gate, for instance as shown in Fig.~\ref{fig:basketball}. The best basis gate would minimize this weighted cost as shown in Eq.~\ref{eq:weighted-cost}.   
\begin{equation}
    {V(U_B)} = \sum_{U_T} f(U_T) D_{U_B}[U_T]
    \label{eq:weighted-cost}
\end{equation}

As gates must typically be calibrated prior to knowing the circuits they will be programmed to implement, a simplified distribution might only consider and 
weight the dominating \cnot{} and \swap{} gates, which, by extension, will generally be true for any \texttt{CPhase} algorithm deployed to the device. We fit the value $\lambda$ as ratio of \cnot{} to the total of \cnot{} and \swap{} gates using our benchmark workloads as illustrated in Fig.~\ref{fig:basketball}. 
\begin{equation}
    W(U_B, \lambda) = \lambda * D_{U_B}[\cnot] + (1-\lambda) * D_{U_B}[\swap]
 \end{equation}
This ratio, $\lambda = 731/(731+828) \approx 0.47$. Therefore, the weighted function $W(U_B, 0.47)$ serves to optimize basis gate selection over quantum workload circuits. We go on to refer to this weighted distribution of gates as $W(\lambda=.47)$.


Table~\ref{tab:gate_counts} compares the decomposition cost of the six common gates from Fig.~\ref{fig:coverage} in terms of number of gates to realize target gates of \swap{} and \cnot{},  as well as Haar and our empirical $W$ distributions. The best performing gate for Haar is the \berk{} because it can span the Weyl chamber in $k=2$, however, \sqiswap{} and \sqberk{} perform well with $k=2.21$ and $k=2.5$, respectively. The $W$ cost requires a gate that is good at both \cnot{} and \swap{}.  While the $K$ function is useful for reasoning about theoretical computational capabilities, the $D$ function better compares the implementation of these gates as it considers pulse times and their impact on the decomposition cost, which we explore in the next section.
\begin{table}[tbp]
\caption{Decomposition Gate Counts ($k$). Each value is determined by the spanning regions from Fig.~\ref{fig:coverage}}
\centering
\begin{tabular}{c||c|c|c|c|c|c}
     &  \iswap & \sqiswap & \cnot & $\sqrt{\cnot}$ & \berk & $\sqrt{\berk}$\\
     \hline
     K[\cnot] & 2 & 2 & 1 & 2 & 2 & 2\\
     K[\swap] & 3 & 3 & 3 & 6 & 2 & 4\\
     $\mathbb{E}$[K[Haar]] & 3 & 2.21 & 3 & 3.54 & 2 & 2.50\\
     K[$W(.47)$] & 2.53 & 2.53 & 2.06 & 4.12 & 2 & 3.06\\
\end{tabular}
\label{tab:gate_counts}
    \vspace{-.2in}
\end{table}

\subsection{Speed-Limit Scaled Duration Costs}
Although each of the discussed candidate basis gates and many others are natively produced by conversion/gain Hamiltonians, different combinations of drives require different duration pulse sequences. 
It follows from Eq.~\ref{eq:conversion-gain} and Eq.~\ref{eq:unitary}, that to realize a specific gate with fixed $\theta_c $ and $\theta_g$, the interaction strengths $g_{c,g}$ are inversely proportional to time $t$. For this reason, a unitary is realizable with the shortest duration when the interaction strengths are as a strong as possible. 

However, in a real physical system, the effective $g_c$ and $g_g$ coefficients cannot be infinitely large due to physical limitations, which can include factors such as fridge heating or disrupting parametric coupling~\cite{zhou2022understanding}.  The maximum magnitude is specific to the system being used. In general, it can be described with a Speed Limit Function (SLF) which describes the valid operating range for variable drive strengths.  The SLF represents the boundary of the regions where the parameters \textit{obey the speed limit} and coupling operates correctly versus where the \textit{speed limit was exceeded} and the unitary gate fails.  

\subsubsection{Characterizing Gate Speed Limits}


 
To illustrate a concrete example of how the speed limit appears in a parametric coupling system and inform our codesign study, we swept the $g_g,g_c$ drive strengths for a Superconducting Nonlinear Asymmetric Inductive eLement (SNAIL) modulator~\cite{frattini20173}.  The gain-, and conversion-only experiments were first performed individually between a qubit and the SNAIL coupler mode to find the maximum $g_c$ when $g_g=0$ and vice versa. This calibrates the relations between the drive amplitudes and the $g$ coefficients. Then, the pumps were  detuned from the on-resonance frequencies (so that the drive affects the SNAIL but we perform no two-qubit  gate) and applied simultaneously to the SNAIL coupler at different amplitude combinations. The result of this study is shown in Fig.~\ref{fig:snail_death}.

To monitor the speed limit, which manifests as a \textit{break point} of the SNAIL coupler, a second qubit that also couples to the SNAIL mode is used. 
This second test qubit is prepared in the ground state and measured immediately after the gain and conversion pumps were turned off. 
Excitation of this second monitoring qubit signals exceeding the speed limit, in which the SNAIL coupler transitions to a (at present poorly understood) chaotic behavior and creates photons in both itself and coupled modes,  illustrated by the red region in the Fig.~\ref{fig:snail_death}.  The blue region indicates the monitoring qubit remained in the ground state and represents our proxy for all the feasible $g_g$ and $g_c$ combinations that can be used to construct 2Q gates.  

As a result, 
the SLF of interest is illustrated as the boundary between the blue coupling region and the red non-coupling region, shown as the white line.    A few characteristics of interest from Fig.~\ref{fig:snail_death}: first, $g_c$ can be driven much harder than $g_g$ and second, the SLF is non-linear.

To capture this experimental information for determining the best basis gate, unitaries  described by the values $g_c$, $g_g$ and time $t$, a gate can be visualized as a line from the origin with the same $g_c$ to $g_g$ ratio that intersects with the SLF to define $g_c^{max}$ and $g_g^{max}$. The ratio of change in drive strength is accompanied by inverse scaling of $t$ to find $t^{min}$.  This process is described in Algorithm~\ref{alg:scale-speeds}.


\begin{algorithm}[bp]
\begin{algorithmic}
    \State \textbf{Input:} \textit{SLF}, $U_B(\theta_c, \theta_g), K_{U_B}[U_T], D[1Q]$
    \State Find the largest $g_c$ and $g_g$ which produces the input U
    \State $\beta \gets \theta_g / \theta_c$
    \State Find intersection of $g_g = \beta g_c$ with \textit{SLF}$(g_c)$ by solving
    $\begin{cases}
    g_g^{max} = \beta g_c^{max}\\
    g_g^{max} = \textit{SLF}(g_c^{max})
    \end{cases}$
    \State Scaling time using updated strengths
    \State $t^{min}\gets \theta_c / g_c^{max}$
    \State Scale decomposition cost by duration
    \State $D_{U_B}[U_T] \gets K_{U_B}[U_T] * t^{min} + (K_{U_B}[U_T] + 1) * D[1Q]$    
    \State \textbf{return} $D_{U_B}[U_T]$
\end{algorithmic}
\caption{Scale Gate Scores using Speed Limit Function}
\label{alg:scale-speeds}
\end{algorithm} 

\begin{figure*}[h]
    \centering
    \subfloat{
        \includegraphics[width=0.3\linewidth]{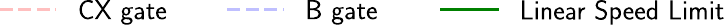}
        \label{fig:slf-best-gate-1Q-0}
    }
    \subfloat{
        \includegraphics[width=0.3\linewidth]{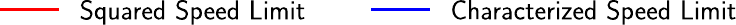}
        \label{fig:slf-best-gate-1Q-0.1}
    }
    \subfloat{
        \includegraphics[width=0.3\linewidth]{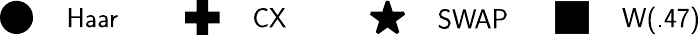}
        \label{fig:slf-best-gate-1Q-0.25}
    }
    \\
    \addtocounter{subfigure}{-3} 
    \subfloat[1Q=0\%]{
        \includegraphics[width=0.16\linewidth]{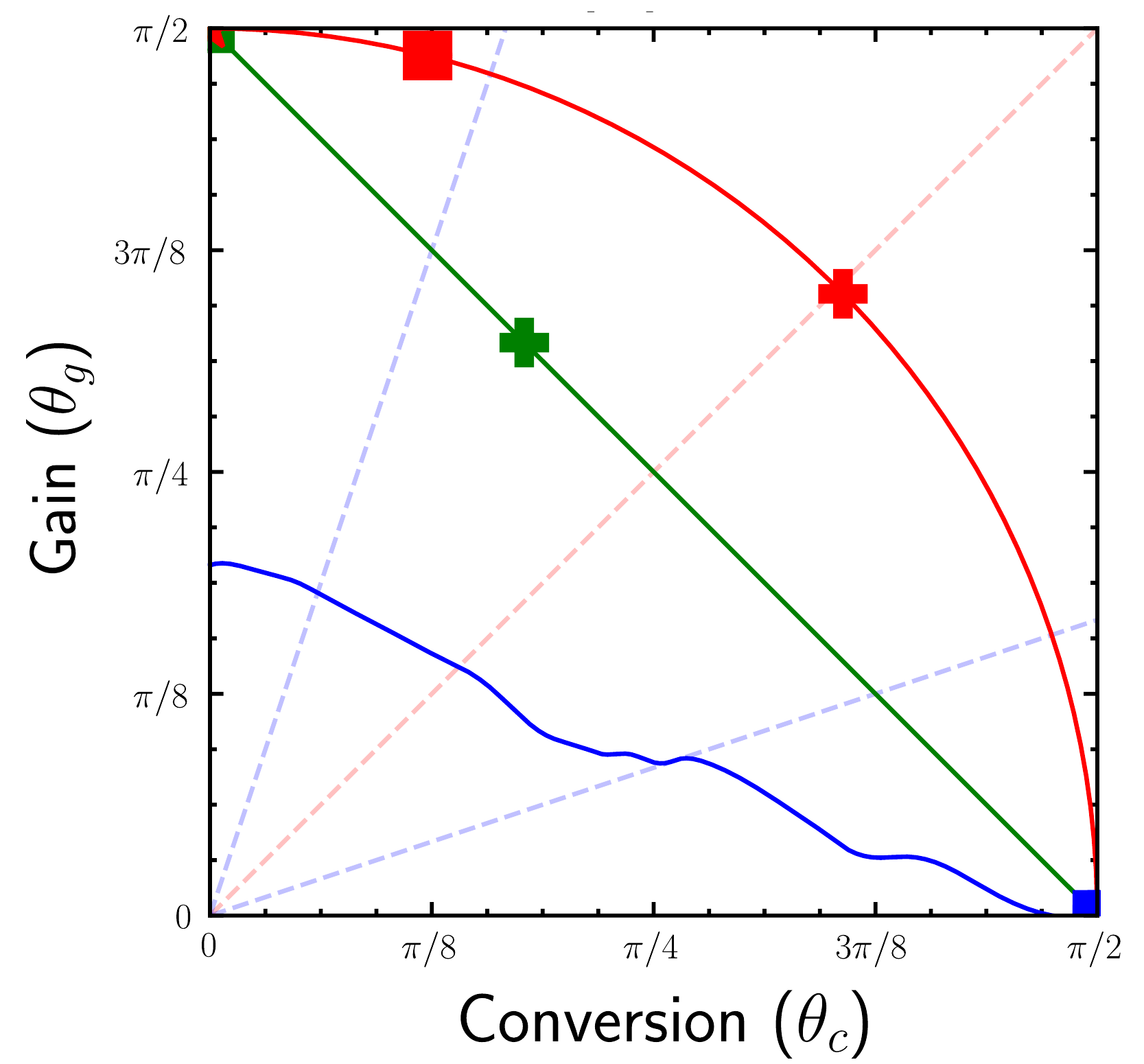}
        \label{fig:slf-weyl-1Q-0}
    }
    \subfloat[1Q=10\%]{
        \includegraphics[width=0.16\linewidth]{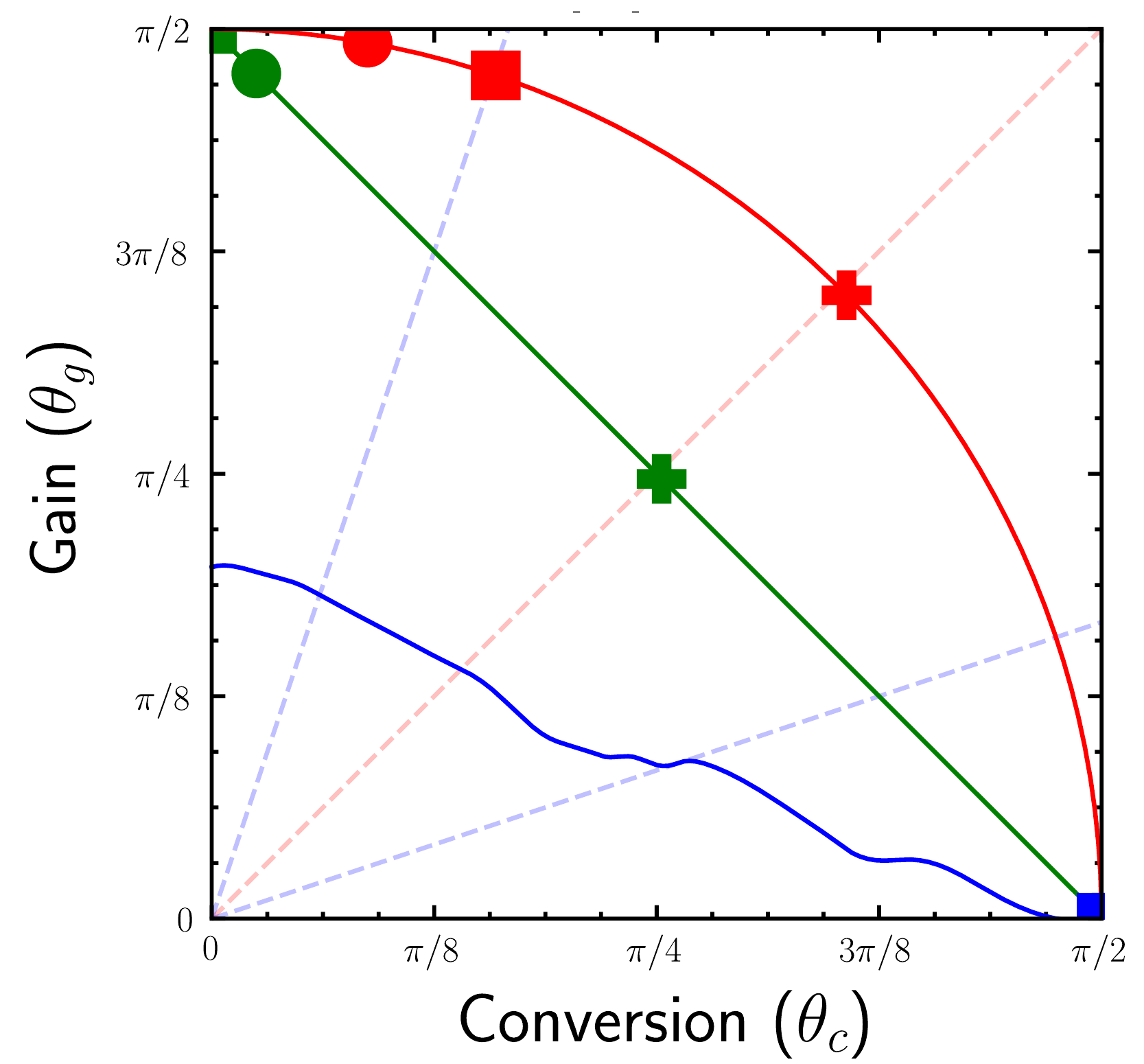}
        \label{fig:slf-weyl-1Q-0.1}
    }
    \subfloat[1Q=25\%]{
        \includegraphics[width=0.16\linewidth]{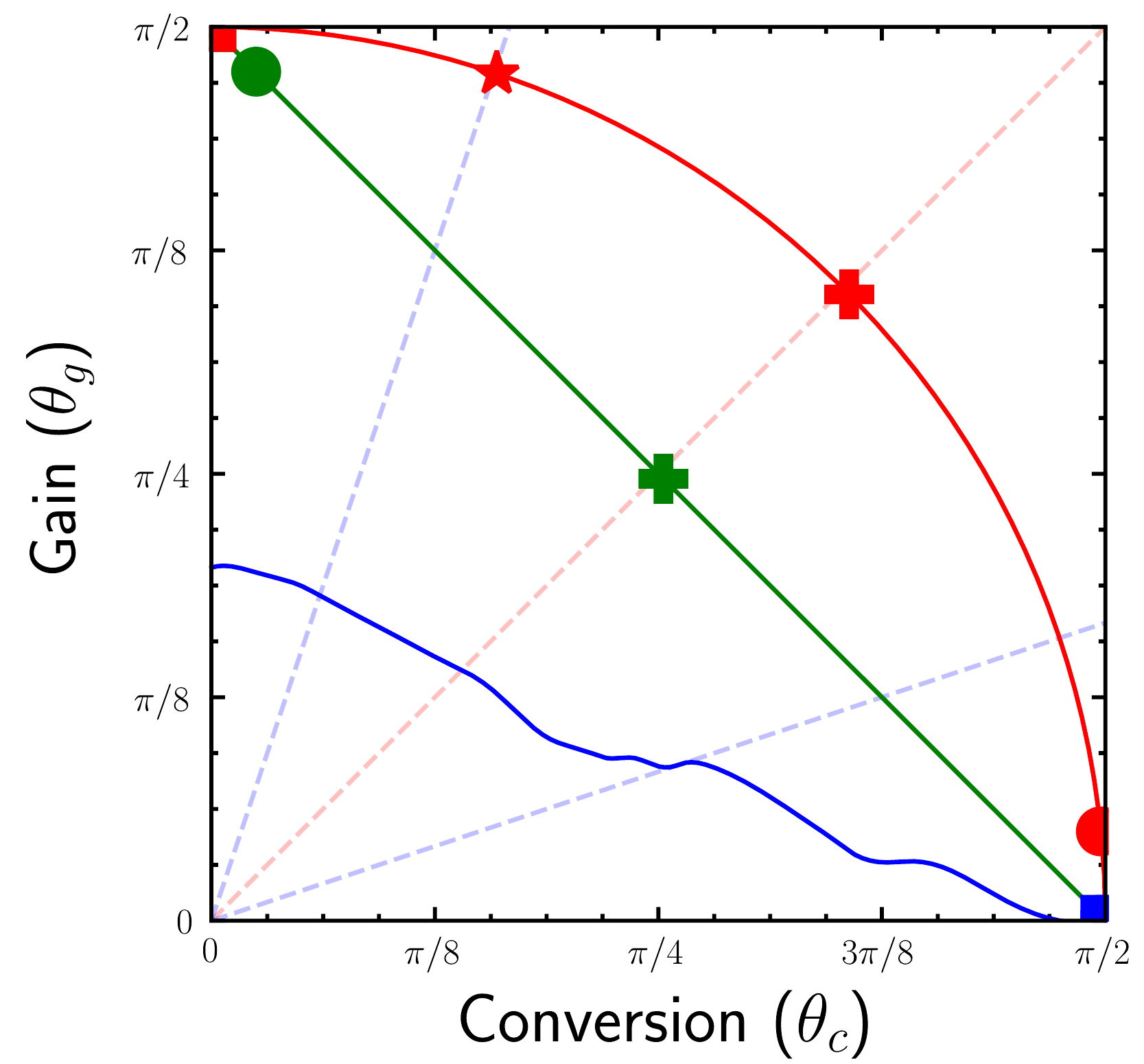}
        \label{fig:slf-weyl-1Q-0.25}
    }
    \subfloat[1Q=0\%]{
    \raisebox{0.15in}{
         \includegraphics[width=0.12\linewidth]{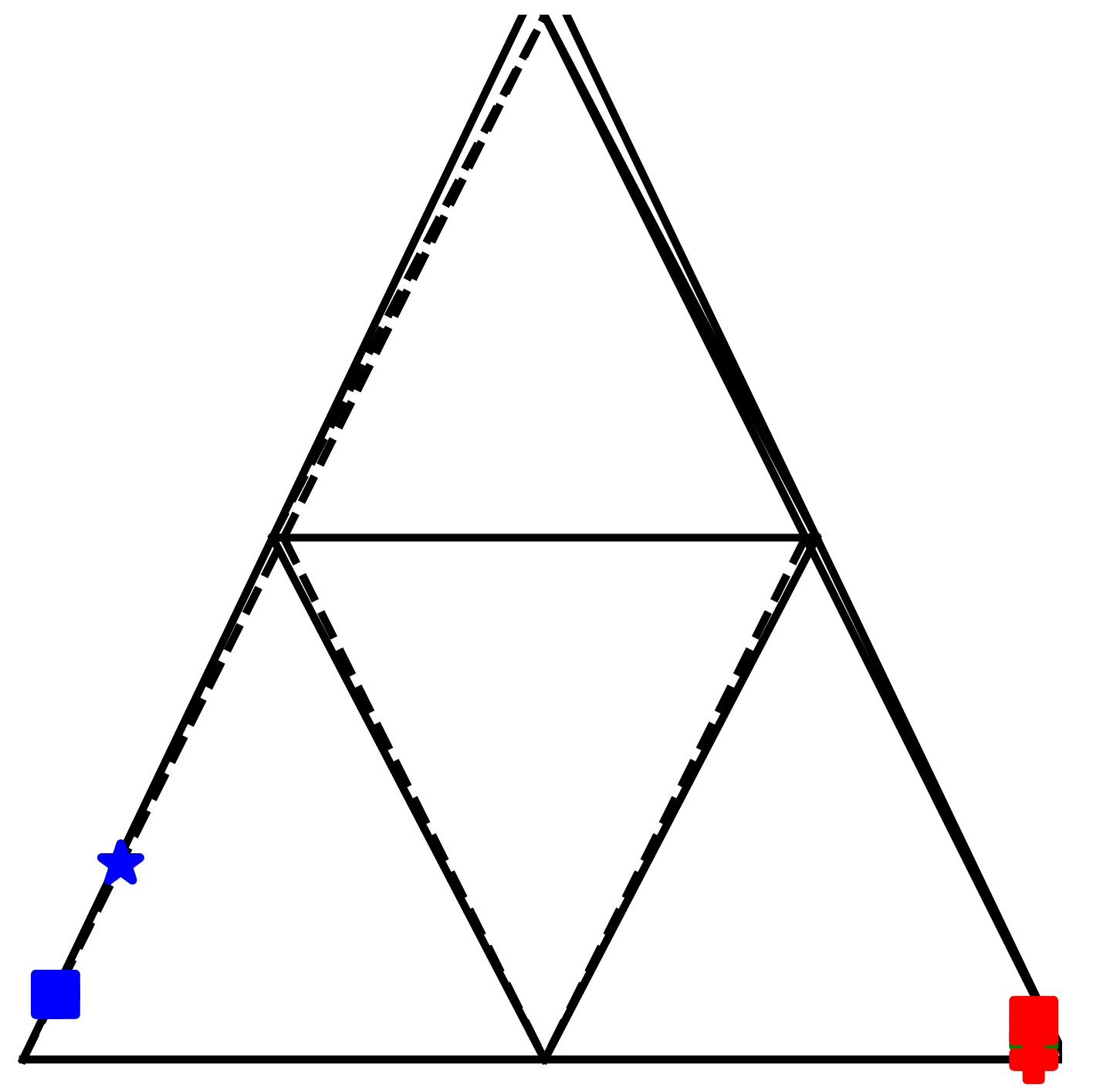}}
        \label{fig:weyl-foo-1Q-0}
    }
    \subfloat[1Q=10\%]{
    \raisebox{0.15in}{
        \includegraphics[width=0.12\linewidth]{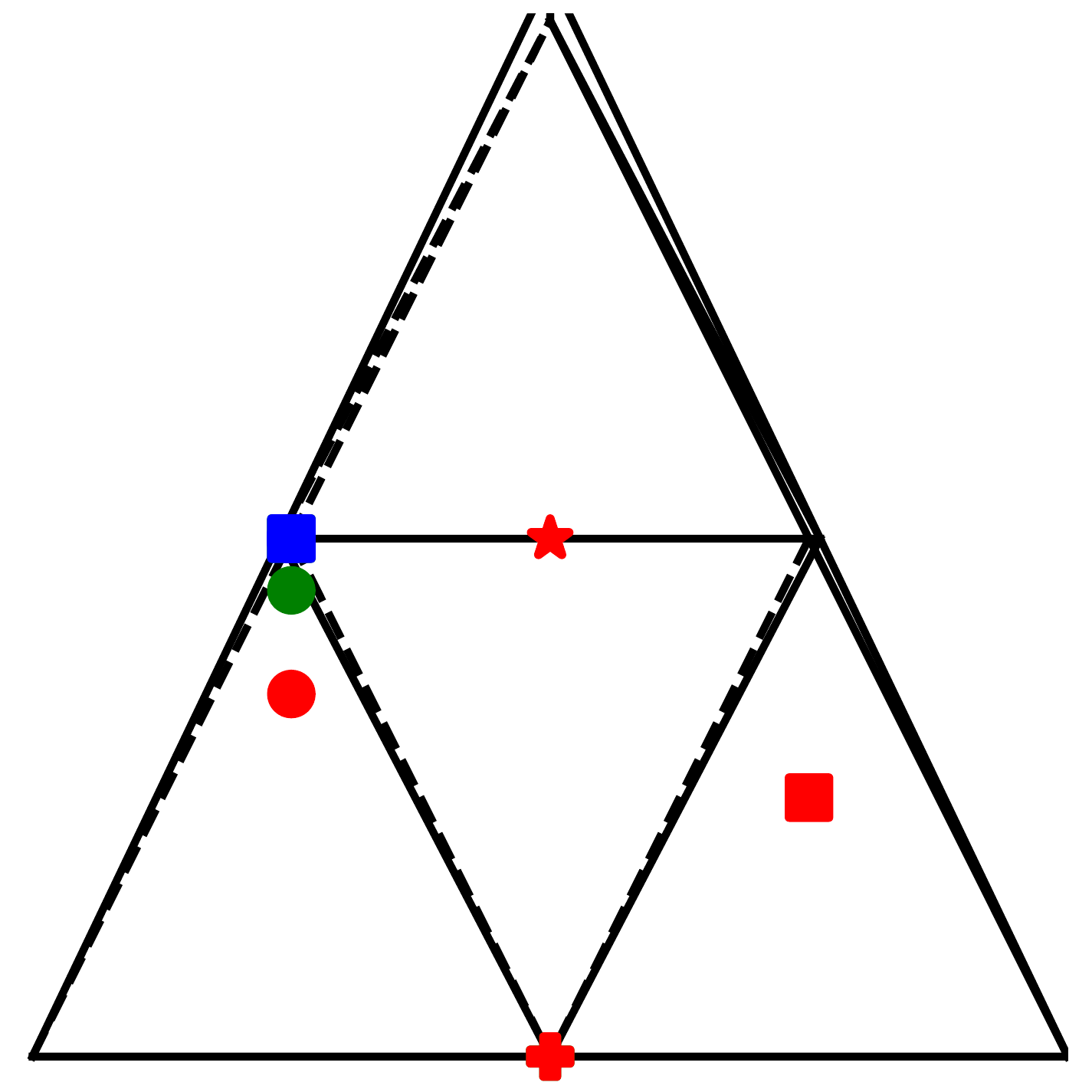}
        \label{fig:weyl-foo-1Q-0.1}
        }
    }
    \subfloat[1Q=25\%]{
    \raisebox{0.15in}{
        \includegraphics[width=0.12\linewidth]{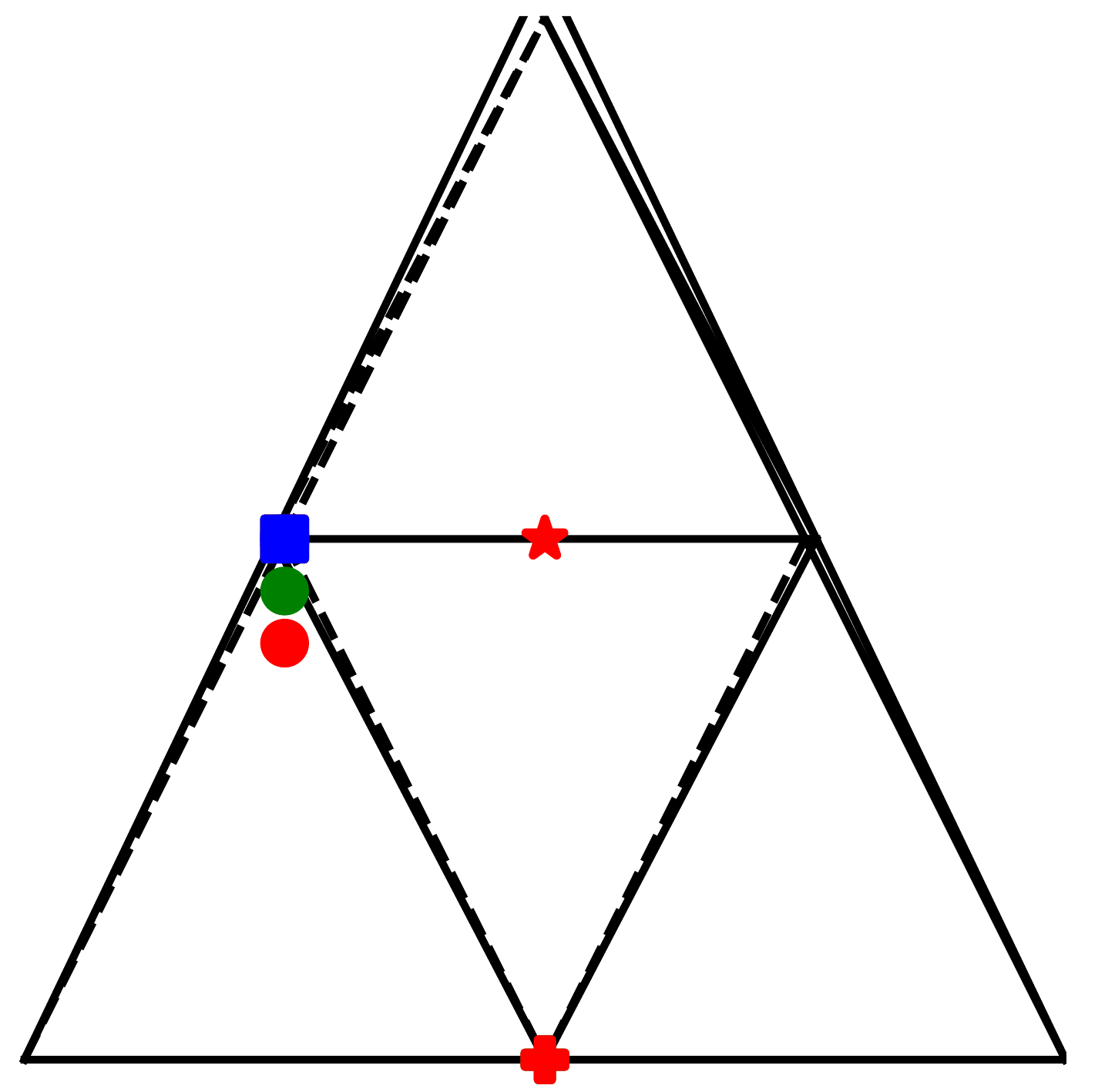}
        \label{fig:weyl-foo-1Q-0.25}
        }
    }\\
    \caption{Best gates for Haar, \cnot{}, \swap{} metrics for varying 1Q gate durations proportional to the length of the basis gate. \colorhl{(a)--(c) represent a continuum of gates using the ratio between $\theta_c$ (x-axis) and $\theta_g$ (y-axis), thus the best gate for some metric is where that line intersects with the defined SLF, scaling each gate to the speed limit. (d)--(f) plot gates in the bottom plane of the Weyl chamber to represent the fractional pulse duration. Gates in the bottom left and right corners, near~$\mathbb{I}$, are smaller fractions of a pulse versus gates closer to the center, which tends towards durations of a single full pulse.}}
    \label{fig:best-gate}
    \vspace{-.1in}
\end{figure*}

To demonstrate the impact of SLFs on basis gate selection we consider two representative synthetic functions in addition to the SNAIL modulator experimental data. The first synthetic function is a linear combination akin to combining voltages, $g_c + g_g \leq L$  resulting in an SLF of $g_g = L - g_c$. The second is a squares of driving strengths function like combining power, enumerated as $g_c^2 + g_g^2 \leq L$ with an SLF of $g_g = \sqrt{L - g_c^2}$. 

To more easily compare these functions, we have normalized the SLF to eliminate any dependencies on hardware specific gate durations. For our synthetic functions, this means selecting $L= \pi/2$. For the SNAIL characterized data we uniformly scale such that the  maximum x- or y-intercept becomes $\pi/2$. In effect, the fastest \iswap{} is normalized to $t^{min}=1$. Now, rather than reporting $D[U_B]$ in units of time, we use units proportional to 1 \iswap{} duration, colloquially referred to as \textit{a single pulse}.  In the next section we explore how these SLFs impact decomposition.

\subsubsection{Circuit Decomposition Costs}

Integrating the SLF into duration efficiency combines the theoretical and practical aspects of gate counts to predict circuit latency. Speed-limited duration of the same popular basis gates reported previously are contained in Table~\ref{tab:basis-durations-linear}. Compared to the theoretical gate counts where \cnot{} and \berk{} both outperformed \sqiswap{}, the speed analysis explains why, in practice, \sqiswap{} becomes the more optimized basis gate, as  \sqiswap{} has lowest consistent pulse cost for Haar score (1.05--1.11) while \sqberk{} has a slightly lower pulse score (0.99) for the squared speed limit.  Moreover, \sqiswap{} also performs well for $W$ (1.27) while \sqberk{} slightly improves on the squared speed limit (1.21).

\begin{table}[tbp]
    \centering
    \caption{Decomposition Duration Efficiency. $\mathrm{D}_{\text{Basis}}$ is the normalized pulse duration for each candidate basis gate based on the SLF. Then, each decomposition score is computed using Table~\ref{tab:gate_counts} and Algorithm~\ref{alg:scale-speeds} over different SLFs.   \textcolor{blue}{Best value reported in blue}, \textcolor{red}{worst value reported in red}.}
    \begin{tabular}{c||c|c|c|c|c|c}
    \hline\hline
         Basis &  \iswap & \sqiswap & \cnot & $\sqrt{\cnot}$ & \berk & $\sqrt{\berk}$\\
         \hline \hline
         
         & \multicolumn{6}{c}{\textbf{Linear Speed Limit}}\\\hline
         $\mathrm{D}_{\text{Basis}}$ & \textbf{1.00} & \textbf{0.50} & \textbf{1.00} & \textbf{0.50} & \textbf{1.00} & \textbf{0.5}\\
         \hline
        
         D[\cnot] & \w{2.00} & \be{1.00} & \be{1.00} & \be{1.00} & \w{2.00} & \be{1.00}\\
         D[\swap] & \w{3.00} & \be{1.50} & \w{3.00} & \w{3.00} & 2.00 & 2.00\\
         $\mathbb{E}$[D[Haar] & \w{3.00} & \be{1.05} & \w{3.00} & 1.77 & 2.00 & 1.25\\
         D[W(.47)] & \w{2.53} & \be{1.27} & 2.06 & 2.06 & 2.00 & 1.53\\
         \hline\hline
         
         & \multicolumn{6}{c}{\textbf{Squared Speed Limit}}\\\hline
         $\mathrm{D}_{\text{Basis}}$ & \textbf{1.00} & \textbf{0.50} & \textbf{0.71} & \textbf{0.35} & \textbf{0.79} & \textbf{0.40}\\
         \hline
         
         D[\cnot] & \w{2.00} & 1.00 & \be{0.71} & \be{0.71} & 1.58 & 0.79\\
         D[\swap] & \w{3.00} & \be{1.50} & 2.12 & 2.12 & 1.58 & 1.58\\
         $\mathbb{E}$[D[Haar] & \w{3.00} & 1.05 & 2.12 & 1.25 & 1.58 & \be{0.99}\\
         D[W(.47)] & \w{2.53} & 1.27 & 1.46 & 1.46 & 1.58 & \be{1.21}\\
         \hline \hline
         
         & \multicolumn{6}{c}{\textbf{SNAIL Characterized Speed Limit}}\\\hline
         $\mathrm{D}_{\text{Basis}}$ & \textbf{1.00} & \textbf{0.50} & \textbf{1.80} & \textbf{0.90} & \textbf{1.40} & \textbf{0.70}\\
         \hline
         
         D[\cnot] & 2.00 & \be{1.00} & 1.78 & 1.78 & \w{2.81} & 1.41\\
         D[\swap] & \w{3.00} & \be{1.50} & 5.35 & 5.35 & 2.81 & 2.81\\
         $\mathbb{E}$[Haar] & 3.00 & \be{1.11} & \w{5.35} & 3.17 & 2.81 & 1.76\\
         D[W(.47)] & 2.53 & \be{1.27} & \w{3.67} & \w{3.67} & 2.81 & 2.15\\\hline\hline
         
    \end{tabular}
    \label{tab:basis-durations-linear}

    \vspace{-.1in}
\end{table}

To find the best basis gate for implementing these target unitaries, the speed limit functions are plotted in Fig.~\ref{fig:best-gate}.  A gate family is defined by the ratio between $g_c$ and $g_g$ terms.  The ratios for the \cnot{} (\texttt{CX}) gate family are shown as blue dotted lines from the origin and the \berk{} gate family is a red dotted line from the origin.  The \iswap{} gate family goes along the x-axis (conversion) and y-axis (gain).

From the figure, the best gate to build a \cnot{} is directly using a \cnot{} basis for the linear and squared SLFs. However, on the characterized system, the \cnot{} is a slow gate, thus it is actually faster to realize an \iswap{} basis and convert into \cnot{} gates. The $W$ function for the squared speed limit is between \iswap{} and \berk{} on the gain side.  For the linear and squared functions the remainder of the gates are at \iswap{} on the gain side. For the SNAIL modulator all gates are pinned at \iswap{} on the conversion side. Recalling Table~\ref{tab:basis-durations-linear}, the basis gate can be from the same gate family but depending on the pulse length, can yield significantly different results.  

We must keep in mind from decomposition rules (Fig.~\ref{fig:decomp-circuit}), templates include interleaved 1Q gates. Our results indicate that for negligible 1Q gate duration, the optimal basis gate is much closer to Identity $\mathbb{I}$, than for appreciable 1Q gates, which tend to be much closer to \sqiswap{}, \cnot{}, and \berk{}. In the next section we discuss the impact of basis gate selection and fractional pulse lengths as impacted by 1Q gates. 
 
\subsection{Interleaving 1Q Gates}

\begin{figure}[tbp]
    \centering
    \includegraphics[width=.75\columnwidth]{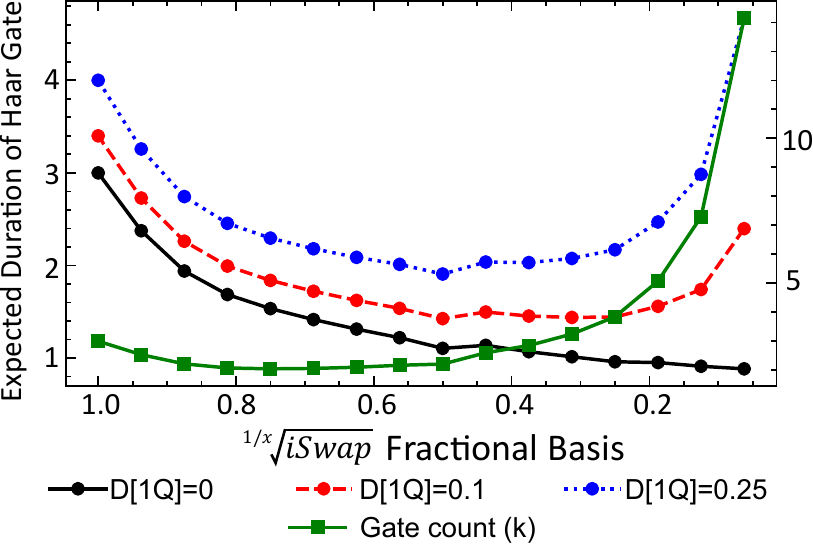}
    \caption{The optimal Haar gate in the iSwap-family changes as function of the 1Q gate times. This is because the number of gate applications increases for smaller cost 2Q basis gates, which trades off with interleaved 1Q layers. \sqiswap{} minimizes duration costs for appreciate 1Q gates.
    }
    \label{fig:haar-fractional-tradeoff}
    \vspace{-.1in}
\end{figure}

\colorhl{
There is a common assumption to treat 1Q gates as negligible, as single qubit interactions are simpler to engineer and as such, less likely to be a significant source of error. Prior work confirms that \sqiswap{} is the more efficient basis gate when only considering 2Q gate costs}~\cite{huang2021towards, mckinney}. 
\colorhl{However, when decoherence is the primary source of error, we find there is an important trade off between faster basis gates and increased $K$-template lengths} (Fig.~\ref{fig:haar-fractional-tradeoff}), thus the accumulated 1Q gate count impacts total duration more for fractional basis gates.

In practice, 1Q gates can be quite fast, \textit{e.g.} around 10\%~\cite{takita2016demonstration} the duration of the basis gate ($D[1Q]=.1$). In other systems with very fast 2Q gates, the 1Q gates are as much as twice as slow as the full pulse 2Q gate~\cite{arute2019quantum,zhao2020high}, depending on the modulator. We treat all 1Q gates as having the same duration, which can be made possible using virtual Z-gates~\cite{mckay2017efficient}.  

When the 1Q gate duration is negligible we prefer to have shorter gates with more repetitions. When the speed limit function is convex, we get a better trade off using \berk{} or \cnot{} gates. The best gates tend towards \sqiswap{} as we observe gates are most efficient when either $g_c$ or $g_g$ is small. Moreover, as shown in Fig.~\ref{fig:best-gate}, when 1Q gate pulse times are 10\% and 25\% of a full pulse, respectively, the best basis gates move \cnot{} to the red dotted line for the linear speed limit and movethe $W$ and \swap{} to the \berk{} family for the squared speed limit at 10\% and 25\%, respectively.  The best Haar score moves out to near, but not exactly to the \iswap{} family. 


When considering 1Q gates, the overall duration of a $U_B$ decomposition can be expressed as in Eq.~\ref{eq:basis-duration}, which sums both the 2Q and 1Q durations for $K$ repetitions. 
To show the impact on decomposition, Table~\ref{tab:asdf} shows the decomposition efficiency for the linear speed limit when 1Q gates are 25\% of the speed of a full pulse 2Q gate.  Similar calculations for other speed limits follow the same trends.
\begin{equation}
    \label{eq:basis-duration}
    D_{U_B}[U_T] = K_{U_B}[U_T] t^{min} + (K_{U_B}[U_T] + 1) D[1Q]
\end{equation}

The total circuit delay can be calculated as in described in Eq.~\ref{eq:full-circuit}, where the pulse delay from Eq.~\ref{eq:basis-duration} is summed for all gates on the critical path of the full circuit.
\begin{equation}
    D_{U_B}[\text{Circuit}] = \sum_{U_T \text{ on Critical Path}} D_{U_B}[U_T]    
\label{eq:full-circuit}
\end{equation}

\begin{table}[tbp]
    \centering
    \caption{Decomposition duration efficiency. Each value is computed using Eq.~\ref{eq:basis-duration}. ($D[1Q] = 0.25$, Linear SLF)}
    \begin{tabular}{c||c|c|c|c|c|c}
         &  \iswap & \sqiswap & \cnot & $\sqrt{\cnot}$ & \berk & $\sqrt{\berk}$\\
         \hline
         D[\cnot] & \w{2.75} & 1.75 & \be{1.50} & 1.75 & \w{2.75} & 1.75\\
         D[\swap] & 4.00 & \be{2.50} & 4.00 & \w{4.75} & 2.75 & 3.25\\
         $\mathbb{E}$[D[Haar] & \w{4.00} & \be{1.91} & \w{4.00}  & 2.91 & 2.75 & 2.13\\         
         D[W(.47)] & \w{3.41} & \be{2.15} & 2.83 & 3.34 & 2.75 & 2.55
    \end{tabular}
    \label{tab:asdf}
    \vspace{-.1in}
\end{table}


Fig.~\ref{fig:haar-fractional-tradeoff} shows that as the length of 1Q gates increases from 10\% of a 2Q gate (red dotted line) to 25\% (blue dotted line), the increasingly small \iswap{} gates reach a practical limit at \sqiswap{}. This motivates revisiting transpiler optimization targeting 1Q gates. The speed limit formulation indicates that the smaller fractional basis gates are advantageous but only if the increased 1Q gate cost can be mitigated. 



\colorhl[]{
Based on the results of this analysis, it is clear that for a linear speed limit, \sqiswap{} is the most duration optimized basis gate. Furthermore, this methodology  offers useful insights for experimentalists when constructing their own basis gates, given their own Hamiltonian design-space and 1Q gate speeds. This is especially pertinent, demonstrated by our hardware speed limit, when there is a strong preference to using one kind of interaction. We also see that as the 1Q gate duration is shorter, the best basis gate approaches Identity, but for appreciable 1Q durations, \sqiswap{} continues to be optimal. In the next section, we introduce parallel-driven gates as a means of improving the basis gate coverage volumes, and consequently reducing duration costs.
}

\section{Parallel 1Q Drive for Basis Optimization}
\label{sec:parallel-drive}


In the previous section, the power of variable conversion/gain drives was explored to show that a wide variety of basis gates could be realized for different pulse lengths with different optimal driving power ratios determined by the speed limit.  However, the same Hamiltonian can be extended if the qubits participating in the 2Q gate are driven simultaneously.  In doing so, it is possible to do at least part of the ``steering" work of the interleaved 1Q layers during the 2Q gate operation in parallel.  This is possible because the drive to the modulator that governs the 2Q interactions is distinct from the drive to the qubits, which would implement the 1Q gates.  An important property of this \textit{parallel-drive} methodology is that it is still only necessary to calibrate a single 2Q gate.

\subsection{Parallel-Driven Hamiltonian}

Transmon Hamiltonians have been explored to optimize pulses for creation of specific gates or algorithms~\cite{kairys2021efficient, GoerzSPP2019, wang2015improving, earnest2021pulse}. In our work, we modify the Conversion-Gain Hamiltonian by appending single-qubit X-gates with drive amplitudes $\epsilon_1(t),\epsilon_2(t)$ each described by $D[2Q]/D[1Q]$ discrete time steps (Eq~\ref{eq:conversion-gain-parallel}). Essentially this creates parallel 1Q gates to occupy the duration of the 2Q gate, each with a distinct amplitude.
\begin{multline}
\hat{H} = g_c (e^{i \phi_c} a^\dag b + e^{-i \phi_c}a b^\dag) + g_g (e^{i \phi_g}ab + e^{-i \phi_g}a^\dag b^\dag) \\ 
+ \epsilon_1(t)(a + a^\dag) + \epsilon_2(t)(b + b^\dag)
\label{eq:conversion-gain-parallel}
\end{multline}

By allowing this extension there are two important outcomes which reduce the overall circuit latency: (1) the basis gate coverage region can be enriched and (2) 1Q gates and their sequential delay may be able to be \textit{absorbed} into the 2Q gate operation, to improve overall circuit time. To consider calibration, 
parallel-drive applies a frequency Kerr-shift on the qubits, such that the parallel-driven gate would require adjustment 
for a qubit frequency shift~\cite{liu2017josephson, zhou2021modular}. Note, the SNAIL is a third-order coupler and minimizes the fourth order Hamiltonian term responsible for cross-Kerr interaction.  This frequency shift is proportional to the fourth-order term, which will not sacrifice qubit fidelity for the SNAIL.  Moreover, it should not impact overhead because existing calibration schemes (\textit{e.g.,} interleaved-randomized and cross-entropy benchmarking) can fine-tune frequency drives of these non-Clifford. Unfortunately, parallel-drive could create additional crosstalk in the IBM cross-resonance gate, however the new IBM initiative to build machines with parametric couplers may also allow high-fidelity parallel-drive.  Next, we examine the impact of parallel-drive on Weyl chamber coverage.

\vspace{-.1in}
\subsection{Computing Parallel Drive Coverage Sets}


\begin{figure}
    \centering
    \includegraphics[width=.7\columnwidth]{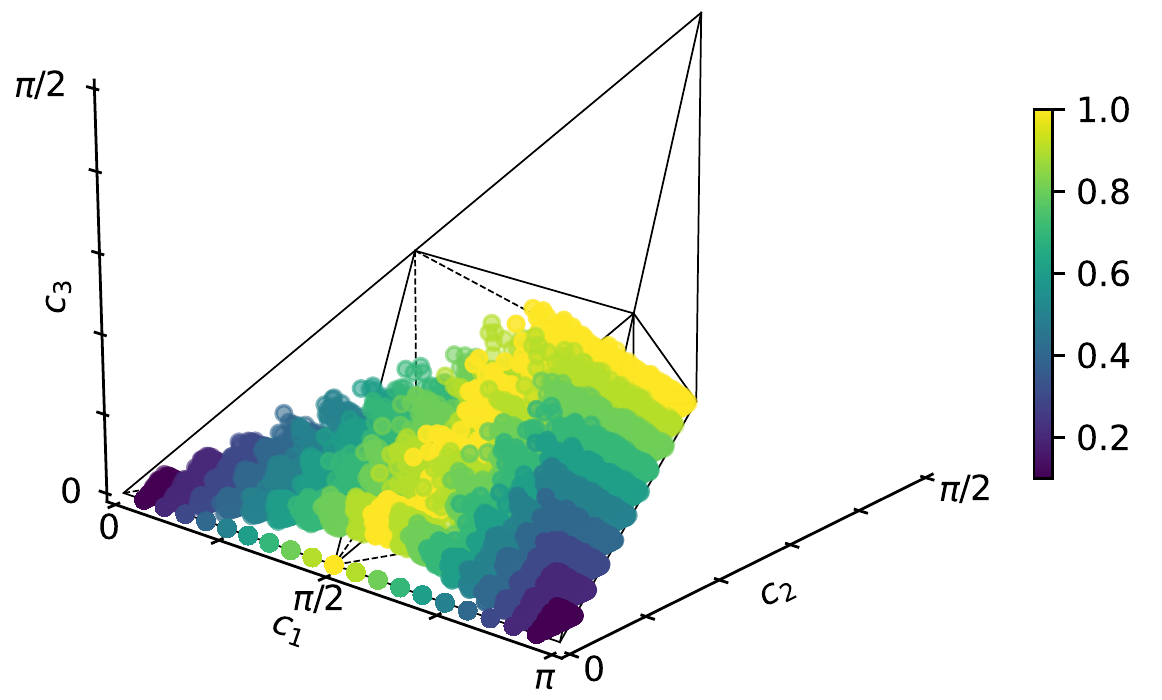}
    \caption{Gates natively produced by conversion and gain parametric driving with parallel 1Q gate drives ($K=1$). The color bar indicates the sum of the $g_c$ and $g_g$, normalized to $\pi/2$.}
    \label{fig:smush-primitives}
    \vspace{-.25in}
\end{figure}

To demonstrate the additional computing capabilities of basis gates using parallel-driven 1Q gates, we first show the increased set of primitive basis gates ($K=1$) found by sweeping the free variables of the Hamiltonian (Eq. \ref{eq:conversion-gain-parallel}), plotted in  Fig.~\ref{fig:smush-primitives} (compared to Fig.~\ref{fig:possible_basis}). The important outcome is that the parallel-driven basis gates extend off the bottom plane into the volume of the Weyl Chamber, which guarantees that our basis templates with parallel-drive, $K'$, will be able to build some targets with fewer iterations than without, $K_0$, \textit{e.g.,} $K'\leq K_0$. This translates to an advantage in Haar Score.

\begin{figure*}
\centering
\subfloat[Decomposition Template]{
    \resizebox{.25\linewidth}{!}{
    \Qcircuit @C=.8em @R=2.0em @!R {
     	& \multigate{1}{\parbox{2.7cm}{2Q $(g_c, g_g, \phi_c, \phi_g, \\ \epsilon_1(t), \epsilon_2(t), T)$}} & \gate{\mathrm{U_{i}}} &\qw & \ustick{K}\\
     	& \ghost{\parbox{2.7cm}{2Q $(g_c, g_g, \phi_c, \phi_g,$} 
      } & \gate{\mathrm{U_{2i}}} & \qw
     	\gategroup{1}{1}{2}{4}{.7em}{\{}
     	\gategroup{1}{1}{2}{4}{.7em}{\}}
     }}
     \vspace{.2in}
    \label{fig:parallel-template}
    }
%
\subfloat[Convergence Plot]{
    \includegraphics[width=.25\linewidth]{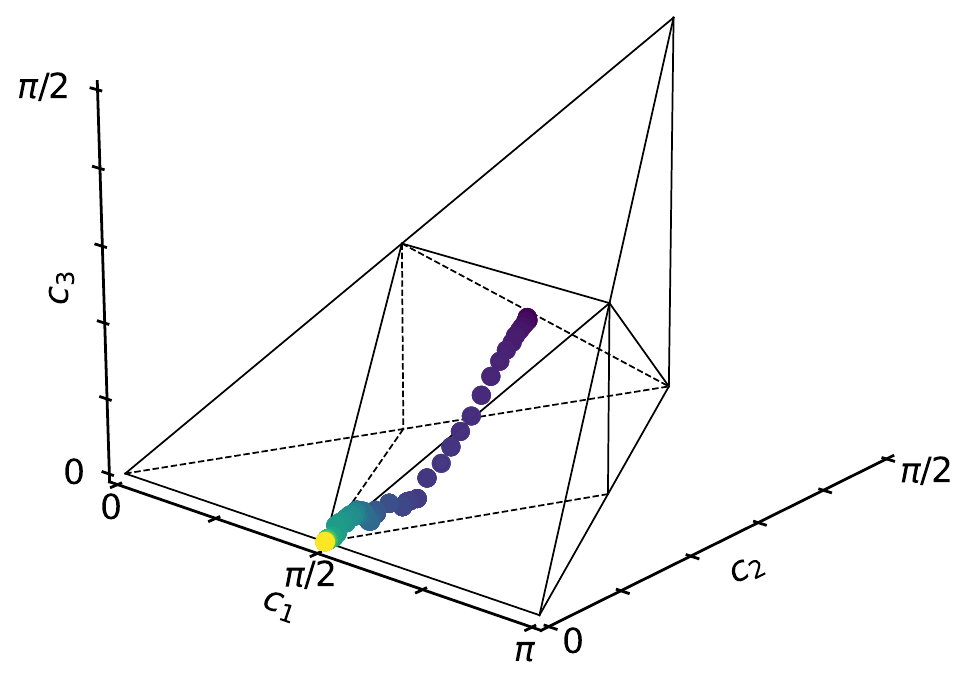}
    \label{fig:iswap_training}}
\subfloat[Iterative Gate Coordinate]{
    \includegraphics[width=.2\linewidth]{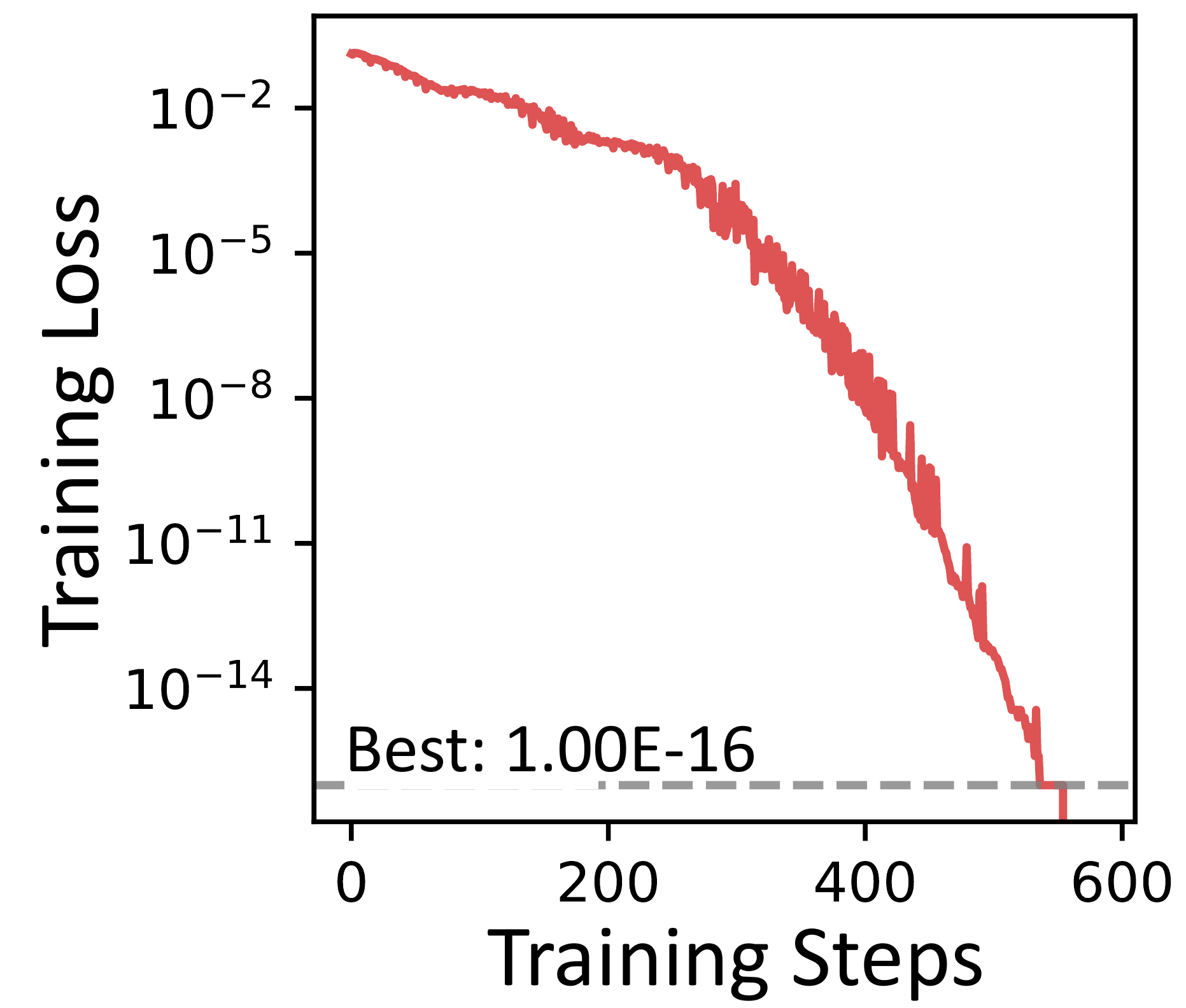}
    \label{fig:iswap_weyl}}
\subfloat[Optimized Cartan Trajectory]{
\includegraphics[width=.25\linewidth]{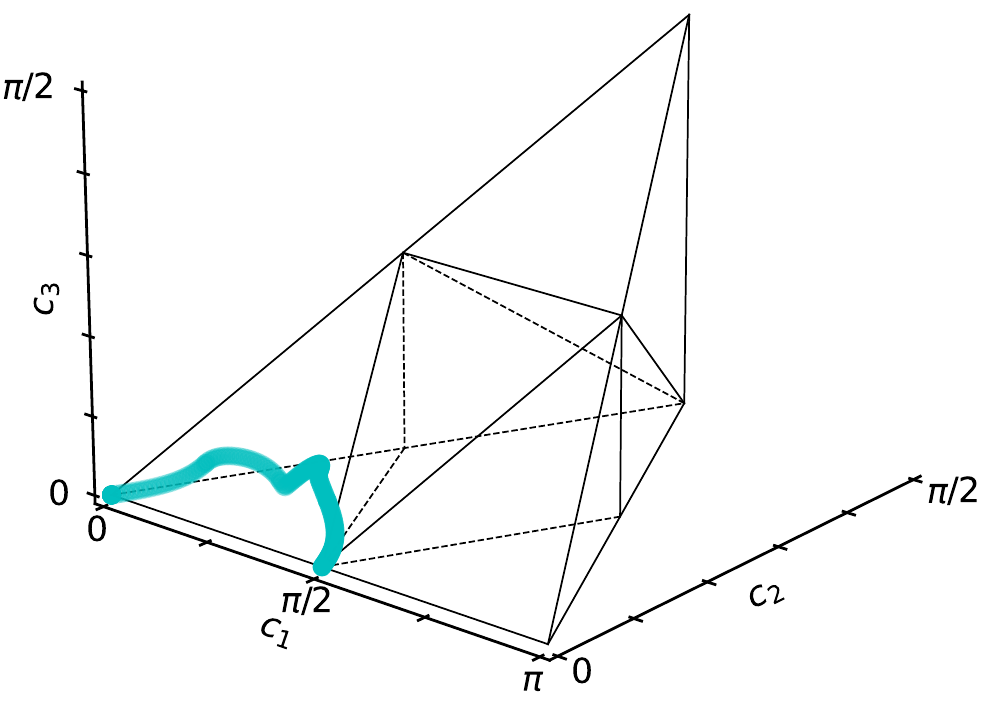}
}

\caption{\colorhl[yellow]{(a) The decomposition template given to the Nelder-Mead optimizer. To bound the coverage regions we attempt to converge to exterior Weyl chamber points. (b)--(c) $K=1$ \iswap{} is verified to contain \cnot{} by optimizing over $\epsilon_1(t)$ and $\epsilon_2(t)$.} (d) The resulting parallel-drive unitary evolution.}
\label{fig:parallel-drive-main}
    \vspace{-.1in}
\end{figure*}

As the analytical volume coverage calculation of monodromy polytopes cannot support 2Q gates with parallel-drive, we developed a numerical procedure to estimate the expanded regions.  First the coverage region is seeded with randomized parameters.  Second, based on the loosely articulated region, target points outside the region are identified to help clarify the outer boundaries.  Using an optimizer to these target points a point at the outside edge of the region can be found.  The approach is described in Algorithm~\ref{alg:parallel-drive-algo}, which constructs a polytope using the $\texttt{lrs}$~\cite{avis1998living} backend via the convex hulls defined by a set of Weyl chamber coordinates. In order to preserve convexity, we partition the coordinate list into left and right sides of the Weyl chamber ($c_1 = \pi/2$), with convex hulls created separately. 
Finally, we specifically target exterior points $\mathbb{I}, \cnot{}, \iswap{}$ and $\swap{}$ as these gates are unlikely to be reached via Haar uniform randomization.

To optimize the template to reach each exterior point, we adapt the strategy from previous work~\cite{nuop, mckinney}, and use the Nelder-Mead optimization method~\cite{nocedal1999numerical} with a Makhlin invariant  functional~\cite{watts2015optimizing, GoerzSPP2019}
. The free variables are phase ($\phi_{c,g}$) and 1Q drive amplitude ($\epsilon_1(t), \epsilon_2(t)$) bounded by $(0, 2\pi)$, for each $K$ iterations of the template, as shown in Fig.~\ref{fig:parallel-template}.



We consider four discrete 1Q drive time steps when building the extended volumes. This corresponds $D[1Q] = .25$, for a full pulse \iswap{}, hence $D[2Q]=1$.  Previous work has explored driving 1Q gates with many more time steps~\cite{kairys2021efficient}; however, in our experimentation, four time steps provides sufficiently similar coverage sets as compared to 250 time steps, but in a more reasonable computing time. For example, Fig.~\ref{fig:iswap_training} plots the norm training loss of an $\iswap{}$ basis converging to \cnot{} in 120 iterations, and Fig.~\ref{fig:iswap_weyl} plots the updated coordinate after each iteration. The final (yellow) coordinate successfully converges to the target \cnot{} at $(\pi/2, 0, 0)$. Note that arbitrarily small error is possible with increased training iterations and 1Q time steps.

\begin{algorithm}[tbp]
\begin{algorithmic}
        \State Basis Template $\gets g_c, g_g, T$
        \State $k \gets 0$
        \While {Coverage Volume not 100\%}
            \State $k \gets k + 1$
            \State Coordinate List $\gets$ []
            \State Randomly Generate Coverage Points
            \For {N iterations}
                \State Template $\gets$ Random($\phi_c, \phi_g,\epsilon_1(t), \epsilon_2(t)$)
                \State U $\gets$ Evaluate(Template)
                \State (x,y,z) $\gets$ Convert U to Weyl coordinate
                \State Coordinate List $\gets$ (x,y,z)
            \EndFor
            \State Train for Exterior Coordinates
            \For {target in ($\texttt{I}, \cnot{}, \swap{}, \iswap{})$}
                \State Save every coordinate along training path
                \State Coordinate List $\gets$ Template.optimize(target)
            \EndFor
            \State Coordinate List partitioned into left and right
            \State Convex Hulls $\gets$ Coordinate List
            \State Basis.Polytope[k]$\gets$ Convex Hulls
        \EndWhile
        \State \textbf{return} Haar Volume(Basis.Polytopes)
\end{algorithmic} 
\caption{Method for Calculating Approximate Improved Volumes from Parallel Drive}
\label{alg:parallel-drive-algo}
\end{algorithm}
\vspace{-.1in}

\subsection{Impact of Parallel Drive on Decomposition}
\begin{figure}[tbp]
\vspace{-.2in}
    \centering
    \subfloat[\iswap]{
    \includegraphics[width=.225\textwidth]{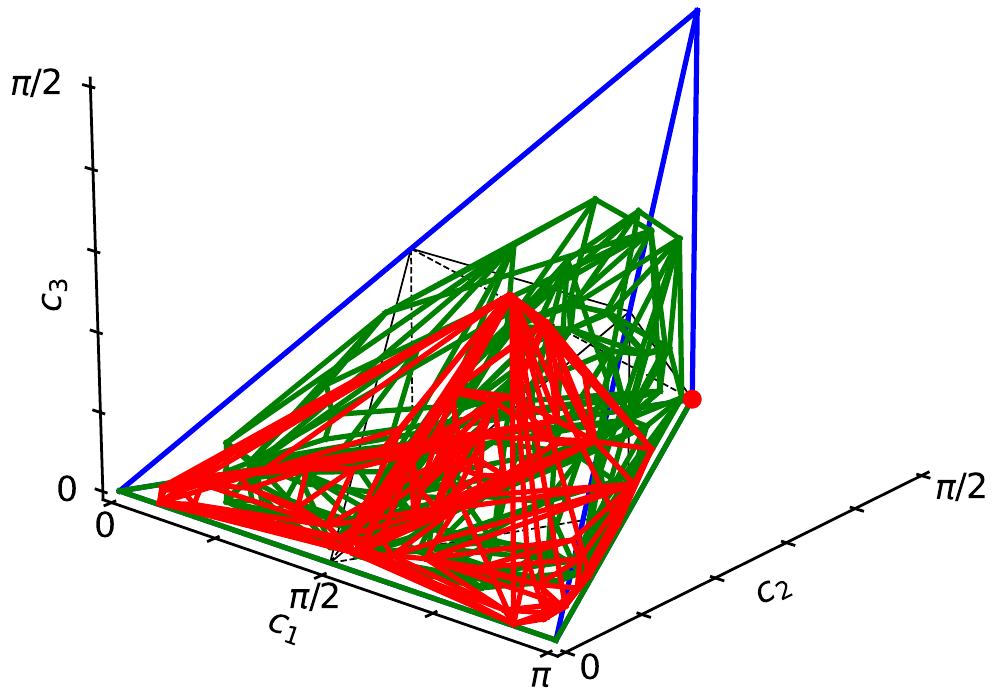}
    \label{fig:iswap_coveragesmush}}
    \subfloat[\sqiswap]{
    \includegraphics[width=.225\textwidth]{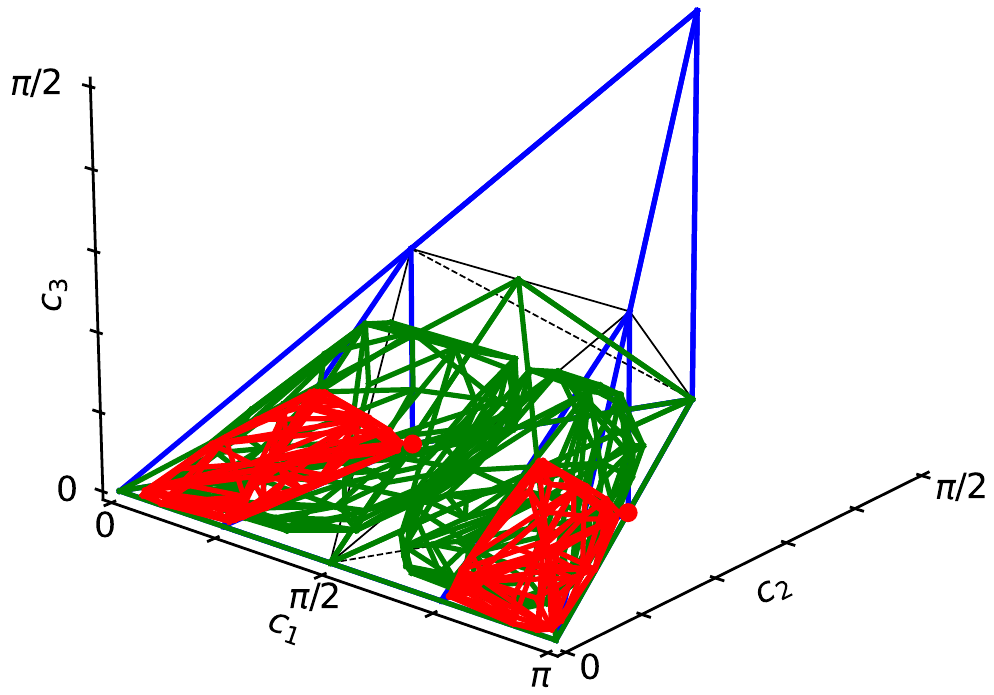}
    \label{fig:sqiswap_coveragesmush}}
    \\
    \subfloat[\cnot]{
    \includegraphics[width=.225\textwidth]{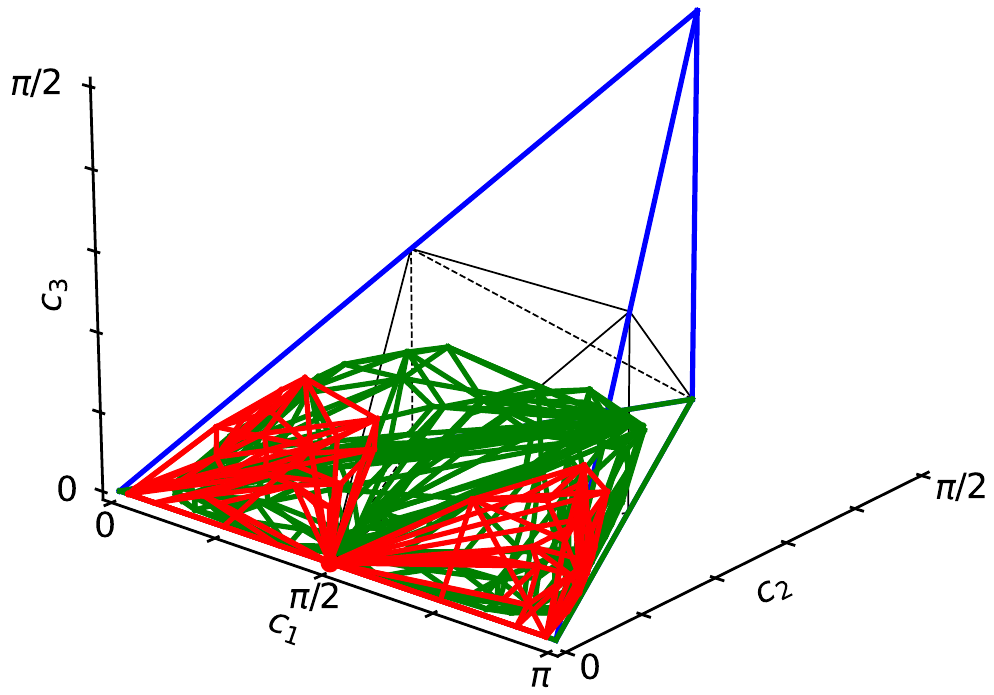}
    \label{fig:cnot_coveragesmush}}
    \subfloat[\sqcnot]{
    \includegraphics[width=.225\textwidth]{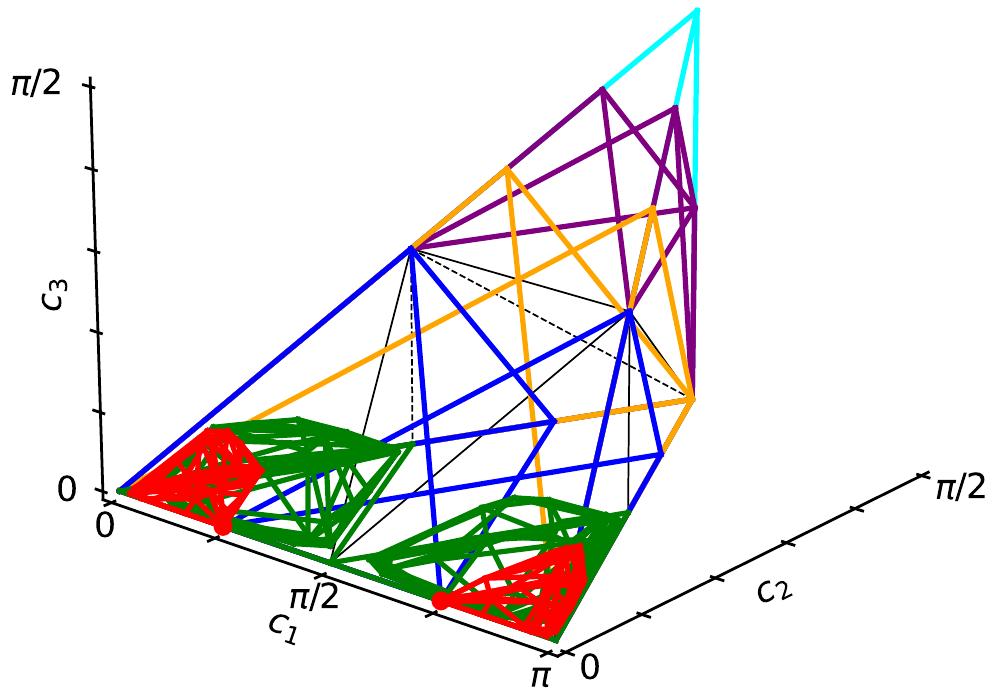}
    \label{fig:sqcnot_coveragesmush}}
    \\
    \subfloat[\berk]{
    \includegraphics[width=.225\textwidth]{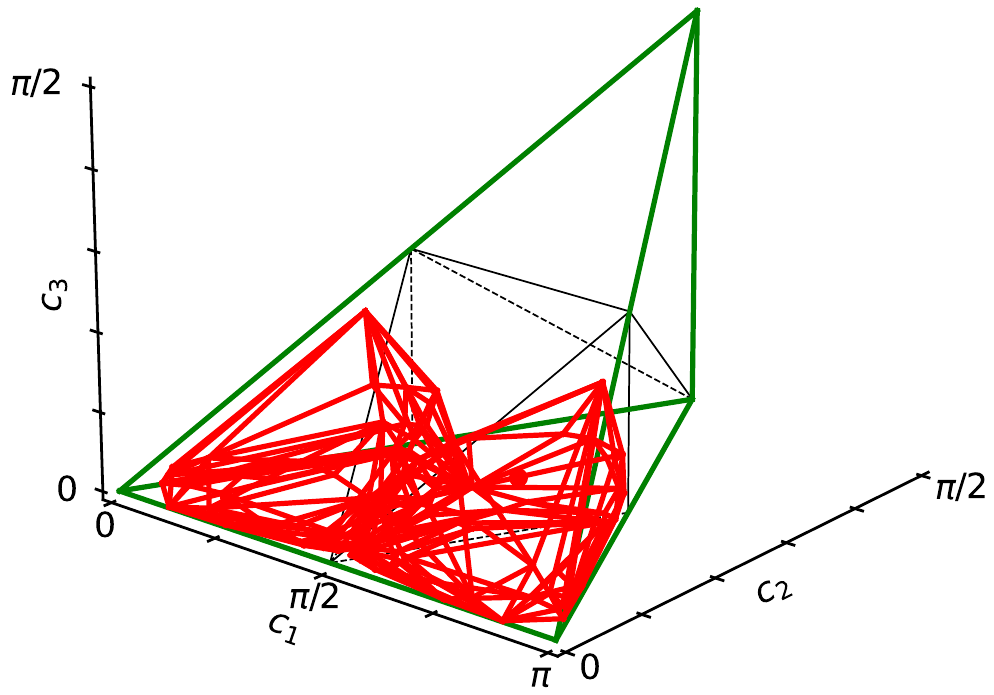}
    \label{fig:b_coveragesmush}}
    \subfloat[\sqberk]{
    \includegraphics[width=.225\textwidth]{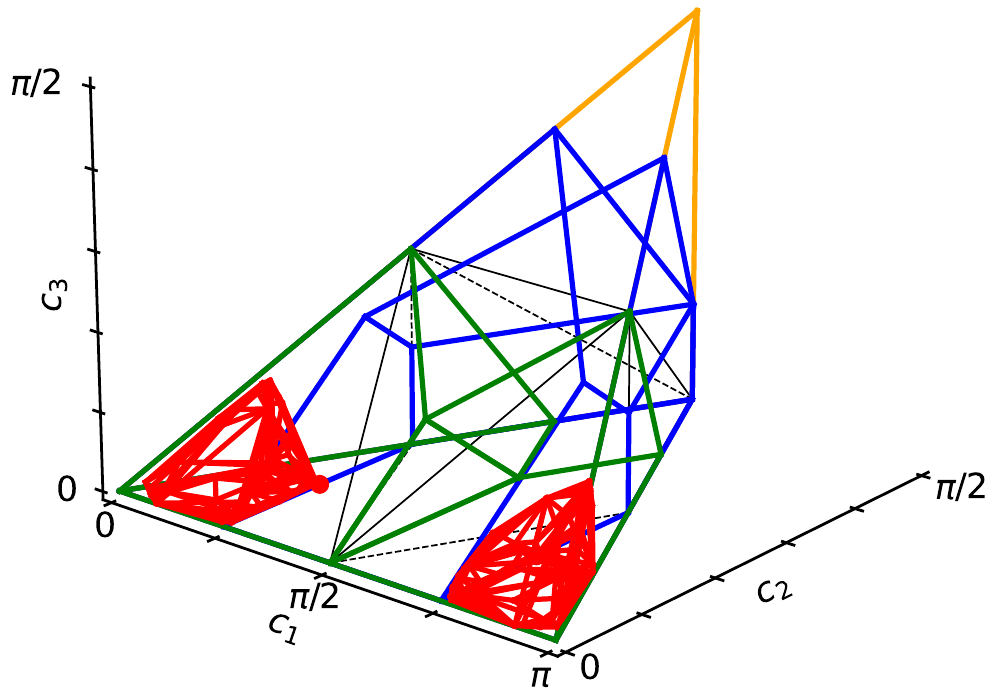}
    \label{fig:sqb_coveragesmush}}
    \caption{Parallel 1Q Drive Extended Gate Coverage Sets. N=3000 random samples. \textcolor{red}{red: $k=1$}, \textcolor{green}{green: $k=2$}, \textcolor{blue}{blue: $k=3$}, \textcolor{orange}{orange: $k=4$}, \textcolor{violet}{purple: $k=5$}, \textcolor{cyan}{cyan: $k=6$} 
    } 
    \label{fig:coverage-smush}
    \vspace{-.1in}
\end{figure}

The extended volumes for each of the six comparative basis gates are reported in Fig.~\ref{fig:coverage-smush}.  The first major difference from the traditional coverage sets (Fig.~\ref{fig:coverage}) is that the $K=1$ in red has increased from being only local to the basis gate into a non-zero volume.  Second, each $K$ spanning region is a superset of its original coverage volumes. Third, no gate reaches 100\% coverage in less template repetitions than before, which highlights the inherent difficulty of optimizing the \swap{} gate, which is always the last gate to be reached.

\begin{table}[tbp]
    \centering
    \caption{Extended basis decomposition gate count cost ($K$). Value is determined by the extended spanning regions in Fig.~\ref{fig:coverage-smush}}
    \begin{tabular}{c||c|c|c|c|c|c}
    \label{table:extended-k}
         &  \iswap & \sqiswap & \cnot & $\sqrt{\cnot}$ & \berk & $\sqrt{B}$\\
         \hline
         K[\cnot] & \be{1} & \w{2} & \be{1} & \w{2} & \be{1} & \w{2}\\
         K[\swap] & \be{2} & 3 & 3 & \w{6} & \be{2} & 4\\
         $\mathbb{E}$[K[Haar]] & \be{1.35} & 2.17 & 2.33 & \w{3.52} & 1.75 & 2.50\\
         K[W(.47)] & \be{1.53} & 2.53 & 2.06 & \w{3.65} & \be{1.53} & 3.06
    \end{tabular}
\vspace{.1in}
    \centering
    \caption{Extended basis decomposition duration cost ($D$) using Parallel-Drive, (D[1Q]=.25, Linear SLF). Fractional basis scores are calculated using the joint spanning regions between themselves and the full basis gate, selecting the lowest cost template.}
    \begin{tabular}{c||c|c|c|c|c|c}
    \label{table:extended-d}
         &  \iswap & \sqiswap & \cnot & $\sqrt{\cnot}$ & \berk & $\sqrt{B}$\\
         \hline
         D[\cnot] & \be{1.5} & \be{1.5} & \be{1.5} & \be{1.5} & \be{1.5} & \be{1.5}\\
         D[\swap] & 2.75 & \be{2.25} & \w{4} & \w{4} & 2.75 & 2.75\\
         $\mathbb{E}$[D[Haar]] & 1.94 & \be{1.71} & \w{3.16} & 2.88 & 2.44 & 2.06\\
         D[W(.47)] & 2.16 & \be{1.90} & \w{2.83} & \w{2.83} & 2.16 & 2.16
    \end{tabular}
    \vspace{-.1in}
\end{table}

Using the same procedure as before, the coverage volumes can be used to find the $K$ and $D$ costs shown in Tables~\ref{table:extended-k} and \ref{table:extended-d}, respectively. 
Note, internal 1Q gates are unnecessary if 
the target gate is identical to multiple fractional copies of the basis gate, \textit{e.g.}, \iswap{} formed from two \sqiswap{}s.  While, this seems trivial for traditional circuits where a straight line in the Weyl chamber continues twice as far, when adding parallel-drive, this property becomes more important as this line becomes a volume in the Weyl chamber, providing more opportunities to eliminate interleaved 1Q gates.  

We use this property to build joint coverage sets between \iswap{} and \sqiswap{} which create decomposition rules using either gate in the decomposition template. 
Interestingly, predominately \sqiswap{} sees a significant advantage from this procedure, as $K=1$ \iswap{} partially covers the perfect entangling region, which is heavily favored by the Haar distribution. For instance, the $K=1$ \iswap{} volume contains the point $(\frac{\pi}{2}, \frac{\pi}{4}, \frac{\pi}{4})$, which renders the $K=2$ \sqiswap{} coverage of the same point unnecessary, and eliminates the duration from the 1Q gate in the decomposition. Both \sqcnot{} and \sqberk{} obey the same rule, but with smaller overlapping volumes. This makes the advantage present, but less significant.
After applying parallel-drive to improve the computing power of each basis gate, we continue to find that \sqiswap{} is the best candidate for a basis gate. Next, we will build explicit decomposition rules into this basis, implement them into a transpilation scheme and report improved simulated fidelities on quantum algorithm benchmarks.

\section{Parallel Drive for \iswap{}-Family}
\vspace{-.1in}

Our work has shown the advantage in calibrating a basis gate with the smallest fraction of total pulse time that does not compromise fidelity.  This can improve Haar score by reducing unnecessary computational work done by longer duration gates.  However, to support calibration, we propose to set the basis gate to approximately the same pulse duration as a 1Q gate.  In the case of the SNAIL modulator, this results in an $\sqrt[4]{\iswap}$ such that $\mathrm{D}[1Q]= 0.25 = \mathrm{D}[2Q]$
Thus the gate can be calibrated without the parallel-drive and again with the parallel-drive to account for the constant frequency shift.
%
%
%
%
This approach minimizes the calibration overhead, as still only a single gate must be calibrated.  However, a \sqiswap{} and \iswap{} can be constructed by two and four $\sqrt[4]{\iswap}$s, respectively.  
Short basis gates are useful for building gates near Identity $\mathbb{I}$, such as the small controlled-phase rotations that appear in QFT; capable of combination to take long strides, \textit{e.g.,} to \swap{}; and can take advantage of parallel-drive to boosts the computational power of gates.

\subsection{Parallel Drive for Improving \cnot{} and \swap{}}
The methodology of creating coverage sets for parallel-driven gates allows us to easily improve decomposition templates via inspection. The \cnot{}-family and \swap{} decomposition rules are given in Fig.~\ref{eq:cx-iswap-template} and Fig.~\ref{eq:swap-iswap-template}. For all other gates, the gate coverage set is used as a lookup table for the required template size. Both the \cnot{} and \swap{} decompositions are shown in Fig.~\ref{fig:cartan-traj}, where the parallel-drive is responsible for the curve in the trajectory. In a full transpilation scheme, the optimizer would be required to fit the exterior 1Q gate parameters, but for the purpose of simulating duration-dependent fidelity, the actual solution is unnecessary.  

\begin{figure}[tbp]
\vspace{-.1in}
\begin{equation*}
\resizebox{.9\columnwidth}{!}{
\Qcircuit @C=1.0em @R=0.4em @!R {
	 	& \multigate{2}{\texttt{CX}(\theta)} & \qw & & & \gate{\mathrm{U}} & \multigate{2}{\texttt{iSwap}(\theta, \epsilon_1(t), \epsilon_2(t))} & \gate{\mathrm{U}} & \qw
	 	\\
	 	& & & \push{\rule{0em}{0em}\!\!\!\! \in \!\!\!\!\rule{0em}{0em}}
	 	\\
	 	& \ghost{\texttt{CX}(\theta)} & \qw & & & \gate{\mathrm{U}} & \ghost{\texttt{iSwap}(\theta, \epsilon_1(t), \epsilon_2(t))} & \gate{\mathrm{U}} &\qw\\
}}
\end{equation*}
\caption{Decomposition template for \cnot{} into \sqiswap{}. A suitable solution for the parallel-drives is $\epsilon_1 =3,\epsilon_2=0$ for all time steps.}
\label{eq:cx-iswap-template}
\vspace{-.2in}
\begin{equation*}
\resizebox{\columnwidth}{!}{
\Qcircuit @C=1.0em @R=0.4em @!R {
	 	& \multigate{2}{\texttt{SWAP}} & \qw & & & \gate{\mathrm{U}} & \multigate{2}{\texttt{iSwap}(\epsilon_1(t), \epsilon_2(t))} & \gate{\mathrm{U}} & \multigate{2}{\sqrt{\texttt{iSwap}}} & \gate{\mathrm{U}} &\qw
	 	\\
	 	& & & \push{\rule{0em}{0em}\!\!\!\! \in \!\!\!\!\rule{0em}{0em}}
	 	\\
	 	& \ghost{\texttt{SWAP}} & \qw & & & \gate{\mathrm{U}} & \ghost{\texttt{iSwap}(\epsilon_1(t), \epsilon_2(t))} & \gate{\mathrm{U}} & \ghost{\sqrt{\texttt{iSwap}}} & \gate{\mathrm{U}}& \qw\\
}}
\end{equation*}
\caption{Decomposition template for SWAP into \sqiswap{}. A suitable solution for the parallel-drives is $\epsilon_1=\pi,\epsilon_2=\pi$ for all time steps. The interior set of 1Q gates is expected to be unnecessary if derived more precisely.}
\label{eq:swap-iswap-template}
 \vspace{-.2in}
\end{figure}

The Weyl chamber does not represent distances and pulse costs uniformly and may mistakenly convey that a shorter, direct trajectory from $\mathbb{I}$ to $U_T$ reduces the required 2Q basis duration, \textit{e.g.} building \cnot{} with parallel-driven \sqiswap{}. However, there is a persistent requirement that 1 total \iswap{} pulse durations appear in the decomposition to be able to reach \cnot{}, likewise 1.5 total \iswap{} pulse duration is required to reach \swap{}. These inherent costs are more rigorously detailed using quantum resource theories~\cite{chitambar2019quantum}, and explain that 2Q decomposition can only be further optimized by removing the 1Q gate delays, but never a shorter 2Q time \textit{i.e.,} the fundamental invariant related to ``computing power." 

\begin{figure}
\centering
\includegraphics[width=.6\columnwidth]{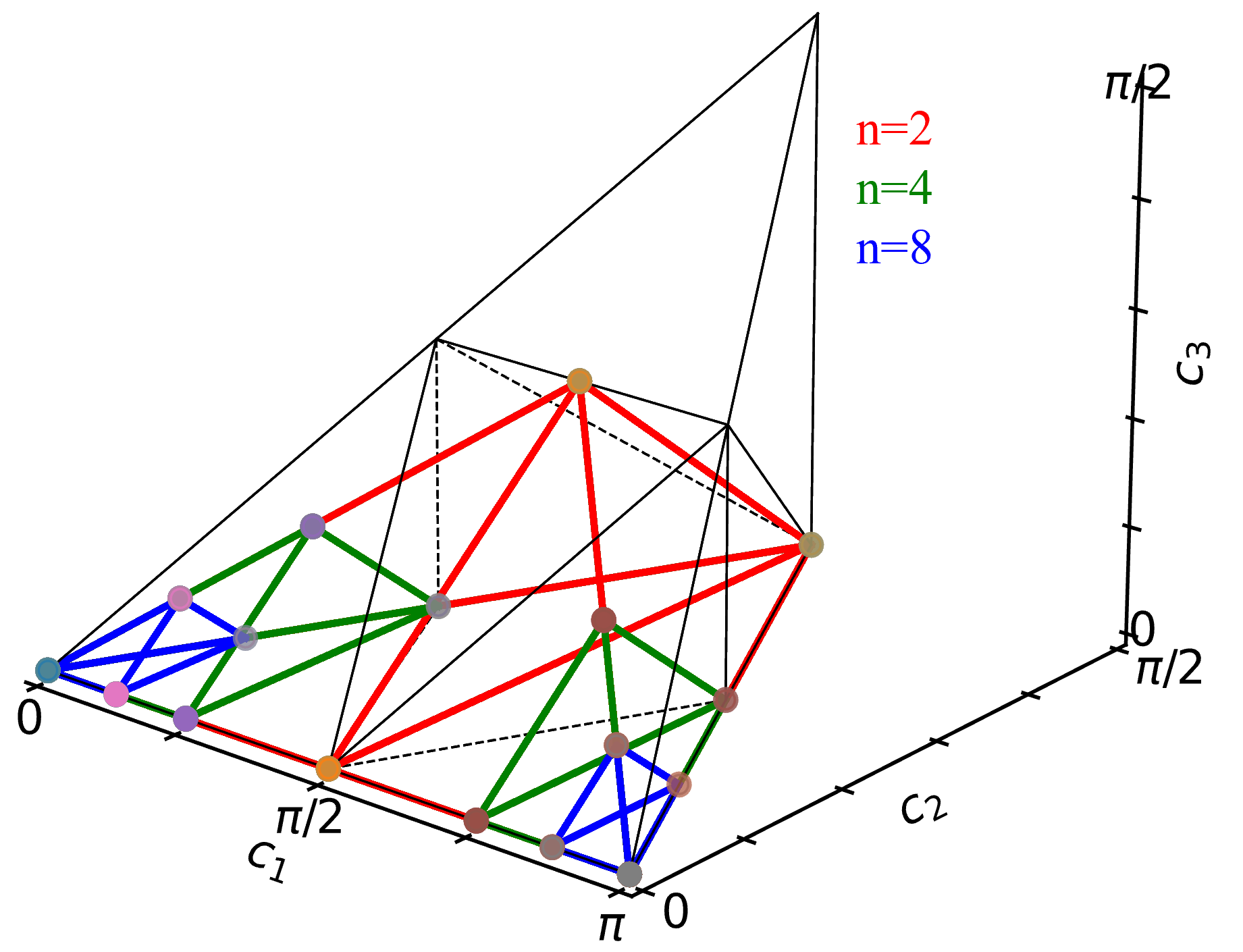}
\caption{Illustrating the $K=2$ coverage of \niswap{n} for $n\in\{2,4,8\}$ which can realize \ncnot{m} for $m\in\{$1,2,4\}.}
\label{fig:iswap-cnot-equal}
    \vspace{-.2in}
\end{figure}

This inherent relationship between \iswap{} and \cnot{} is depicted in Fig.~\ref{fig:iswap-cnot-equal}, such that a fractional duration \iswap{} always contains the same fractional duration \cnot{}. For instance, K=2 with \sqiswap{} or K=1 with parallel-drive \iswap{} both reach \cnot{}. Of course, \sqiswap{} is still the more powerful basis despite this relation to \cnot{} as it always contains additional Weyl Chamber coverage. Both \iswap{} and \cnot{} are special perfect entanglers, and interestingly a non-entangling \swap{} gate can be used to convert back and forth between gates~\cite{crooks2020gates}.

\subsection{Simulated Fidelity Improvements}

We utilize a circuit fidelity model that captures the primary source of error as decoherence in time following the methodology of prior work~\cite{gokhale2021faster,mckinney}. The fidelity of a final qubit state, $\mathcal{F}_Q$, over a single path or wire in the circuit exponentially decays as a function of the ratio of circuit duration time and the qubit's $T_1$ (relaxation rate). Then the total circuit fidelity, $\mathcal{F}_T$ is given by the composite final qubit state, and thus is exponential with number of qubits. For this reason, even small improvements in circuit duration cascade into improved path and total circuit fidelities. 
\begin{align}
    \mathcal{F}_Q &= e^{-D[\text{Circuit}]/T_1}\\
    \mathcal{F}_T &= \prod_{i=1}^N {\mathcal{F}_{Q_i}}
\end{align}
%
%

Our transpilation scheme uses the SLF normalized durations $D[U_B]$, which are converted back into units of time by multiplying by the \iswap{} duration. 
To quantify these improvements we choose $D[\iswap{}] = $~\SI{100}{\nano \second} 
and $D[1Q] = $~\SI{25}{\nano \second} 
with qubit lifetime $T_1= $~\SI{100}{\micro \second}
, which is consistent with transmon qubits using a SNAIL modulator~\cite{zhou2021modular}. Using these values, the improvements from the reduced duration decompositions over \cnot{}, \swap{}, and Haar random targets are given in Table~\ref{tab:gate_improvements}. The baseline uses previously derived analytical \sqiswap{} decomposition rules~\cite{huang2021towards}. Note that although exterior gates are used in the \cnot{} decomposition to make it perfectly equivalent, quantum algorithms often have 1Q gates before and after \cnot{} gates. Therefore, the circuit's 1Q gates and the decomposition substitution's 1Q gates would naturally combine for an even lower cost than represented here.

In our transpilation flow, we start by consolidating runs of all unitary blocks into 2Q gates and inducing \swap{}s on a $4\times4$ square-lattice topology\footnote{A \cnot{} followed by a \swap{} on the same qubit pair is equivalent to an \iswap{} which appears with non-negligible frequency, see Fig.~\ref{fig:basketball} at \iswap{}.}. We then decompose each gate into the \sqiswap{} basis with the pulse duration calculated by the provided SLF. 
Decomposition uses predefined substitutions for gates locally equivalent to \cnot{}-family and \swap{} gates (see Fig.~\ref{fig:parallel-drive}). If a rule is not known, we load the \iswap{} and \sqiswap{} extended coverage sets to construct a minimum size $K$ template. Finally, we consolidate consecutive 1Q gates and report the remaining durations on each path. 

\begin{table}[tbp]
    \centering
    \caption{Improved gate infidelities, $1-\mathcal{F}_Q$ (D[1Q]=.25, Linear SLF)}
    \begin{tabular}{c||c|c|c}
    $U_T$ & Baseline & Optimized & \% Improved \\
    \hline
    
    \cnot{} & 0.0035 & 0.0030 & 14.3\\
    \swap{} & 0.0050 & 0.0045 & 9.98 \\
    \ehaar{} & 0.0038 & 0.0034 & 10.5\\
    W(.47) & 0.0043 & 0.0038 & 11.62
    
    \end{tabular}
    \label{tab:gate_improvements}
    \vspace{0.1in}
    \centering
    \caption{Transpilation Results (D[1Q]=.25, Linear SLF). Baseline and Optimized columns report total circuit duration in D[2Q]=1 normalized units. Duration, $\mathcal{F}_Q$, and $\mathcal{F}_T$ columns are reported as the relative \% improvement between the baseline and optimized durations.}
    \begin{tabular}{c||c|c|c|c|c}
        Benchmark & Baseline & Optimized & Duration & $\mathcal{F}_Q$ & $\mathcal{F}_T$\\
        \hline
        QV & 133.0 & 118.4 & 11.22 & 1.50 & 27.0 \\
        VQE\_L & 25.75 & 21.5 & 16.50 & 0.43 & 7.04 \\
        GHZ & 31.75 & 27.00 & 14.96 & 0.48 & 7.90 \\
        HLF & 102.3 & 88.00 & 13.94 & 1.43 & 25.6 \\
        QFT & 149.5 & 120.3 & 19.53 & 2.96 & 59.5 \\
        Adder & 175.0 & 144.3 & 17.57 & 3.12 & 63.6\\
        QAOA & 197.8 & 147.8 & 25.25 & 5.12 & 122\\
        VQE\_F & 333.3 & 286.8 & 13.95 & 4.76 & 110\\
        Multiplier & 1065.25 & 770.76 & 27.64 & 34.2 & 11000
    \end{tabular}
    \label{tab:final-results}
    \vspace{-.1in}
\end{table}


Our decomposition improvements using parallel-drive led to an average relative reduction in duration of 17.8\% for the set of quantum algorithm workloads, as determined by selecting the best outcome from 10 transpiler runs, reported in Table~\ref{tab:final-results}.
The Quantum Volume results were further averaged through additional runs due to the random nature of the algorithm.

The average relative reduction in duration is directly related to our improvement method, while the relative path and total fidelity improvements depend on the baseline duration, mock gate durations, and qubit lifetime. Shallower circuits inherently have higher fidelities, and thus, their improvement potential is limited compared to deeper circuits with lower fidelities. For instance, the Quantum Volume improves from 0.875 to 0.889 in terms of path fidelities (a 1.5\% improvement), which, due to its exponential relationship in the number of qubits, results in an increase from 0.119 to 0.151 (a 21\% improvement) in total fidelity. In contrast, the shortest VQE\_L algorithm path fidelities baseline of 0.975 only improve to 0.979 (0.4\% improvement), leading to a total fidelity increase from 0.662 to 0.709 (6.6\% improvement). Finally, our W(.47) metric predicts an average 11.6\% reduction in duration. Our experimentally demonstrated 17.8\% actually outperforms this case due to additional improvements to \cnot{} where the decomposition template's exterior 1Q gates could be merged or eliminated.

\section{Conclusion}

In this paper, we formally characterized the optimal basis gate for a parametric coupler under hardware speed limitations. The results indicate that, despite the \sqiswap{} being close to optimal prior to our analysis, it can still be improved by utilizing parallel 1Q gates. This small improvement leads to a notable enhancement in fidelity as the number of qubits increases. Our co-design evaluated uniform Haar gates and circuit-based gate sets, finding that for realistic cost functions such as our experimentally-determined SNAIL-coupler data, the \sqiswap{} gate performed the best in nearly all scenarios. 

Initially, gate count scores favored the \berk{} gate, but after considering the cost of direct generation through multiple simultaneous parametric drives, the \sqiswap{} gate was the most efficient.
The introduction of parallel-drive and related transpilation optimizations reduced the gate duration for most basis gates and improved the pulse time for the \sqiswap{} gate 
The \iswap{} family was uniquely enhanced through joint parallel-drive extended coverage sets, yielding significant improvements in fidelity due to faster circuit execution.  

In future work, we aim to expand our parallel-drive transpilation flow to further enhance compilation strategies for quantum algorithms and test them on various quantum systems with differing speed limit characterizations and dynamics.  
More, detailed studies of improvement of parallel-drive volume versus calibration complexity for different quantum machine targets, including studying calibration complexity, while expanding the 
flexibility to handle continuously variable drive parameters, similarly to optimal-control theory methods are important next steps.

\section{Acknowledgements}
This work is supported by the University of Pittsburgh via a SEEDER grant, by the DOE via the C2QA collaboration, and NSF Award CNS-1822085. CZ, MX, and MJH are partially supported by the U.S. Department of Energy, Office of Science, National Quantum Information Science Research Centers Co-Design Center for Quantum Advantage under contract DE-SC0012704.

\section{Supplementary Material}
The code used in this study is available at \url{https://github.com/Pitt-JonesLab/slam_decomposition}. This repository includes the scripts to create circuit templates for decomposition, calculate parallel-driven expanded coverage sets, and implement the transpiler.

\balance
\bibliographystyle{IEEEtran}
\bibliography{refs}

\end{document}